\documentclass[11pt]{article}
\pdfoutput=1
\usepackage{jheppub}
\usepackage{amsmath}
\usepackage{bm}
\usepackage{slashed}
\usepackage{graphicx}

\newlength{\dummysp}
\newcommand{\half}{\frac{1}{2}}

\newcommand{\beq}{\begin{eqnarray}}
\newcommand{\eeq}{\end{eqnarray}}

\newcommand{\gappeq}{\mathrel{\rlap {\raise.5ex\hbox{$>$}}
{\lower.5ex\hbox{$\sim$}}}}
\newcommand{\lappeq}{\mathrel{\rlap{\raise.5ex\hbox{$<$}}
{\lower.5ex\hbox{$\sim$}}}}

\newcommand{\ben}{\begin{enumerate}}
\newcommand{\een}{\end{enumerate}}

\newcommand{\bit}{\begin{itemize}}
\newcommand{\eit}{\end{itemize}}

\def\[{\left [}
\def\]{\right ]}
\def\({\left (}
\def\){\right )}
\def\R{{\mathbb R}}

\def\Z{{\mathbb Z}}
\def\circle{{\mathbb R^3\times S^1_L}}
\def\torus{{\mathbb R^2\times S^1_L\times S^1_{\beta}}}
\pdfoutput=1

\title{Deconfinement on $\mathbb R^2\times S^1_L\times S^1_{\beta}$ for all gauge groups and duality to double Coulomb gas}

\author[]{Brett Teeple}  
 
\affiliation[]{Department of Physics,   University of Toronto, 
Toronto, ON M5S 1A7, Canada}
\emailAdd{bteeple@physics.utoronto.ca}   
    
\abstract
{I study finite-temperature $\mathcal N=1$ super Yang-Mills for any gauge group $G=A_N, B_N, C_N, D_N, E_{6,7,8},F_4,G_2$, compactified from four dimensions on a torus, $\mathbb R^2\times S^1_L\times S^1_{\beta}$. I examine in particular the low temperature regime $L\ll\beta=1/T$, where $L$ is the length of the spatial circle with periodic boundary conditions and with anti-periodic boundary conditions for the adjoint gauginos along the thermal cycle $S^1_{\beta}$. For small such $L$ we are in a regime were semiclassical calculations can be performed and a transition occurs at $T_c$ much smaller than $1/NL$. The transition is mediated by the competition between non-perturbative objects including 'exotic' topological molecules: neutral and magnetic bions composed of BPS and KK monopole constituents, with $r=rank(G)$ different charges in the co-root lattice of the gauge group $G$, and the perturbative electrically charged W-bosons (along with their wino superpartners).\\
The difference from non-SUSY theories here is that the Higgsing along the thermal cycle gives rise to a light modulus scalar field which couples to both bion-instantons and the W-bosons, and mediates a transition near $T_c$ where the bions and W-bosons compete with equal strengths. The transition is seen to be similar to previous studies on $\mathbb R^3\times S^1_L$ [1,2,12,13] with general gauge group where a first order transition was found for all groups, but a second order one for the case of $SU(2)$ on the torus $\mathbb R^2\times S^1_L\times S^1_{\beta}$, which was subjected to lattice studies in [1].\\
I determine a duality to a double Coulomb gas of neutral and magnetic bions of different charges of their constituent monopole-instantons, and W-bosons of both scalar and electric charges. Aharanov-Bohm interactions exist between magnetic bions and W-bosons, and scalar charges of W-bosons and neutral bions attract like charges, as opposed to the magnetic and electric charges where like charges repel. It is hoped in the future that lattice studies of this Coulomb gas can be done as in [1] for all gauge groups. It is hoped that a dual lattice 'affine' XY model with symmetry breaking perturbations can also be found in future studies of general gauge group as done in [1] for $SU(2)$.}

\begin{document}
\maketitle
\flushbottom

\section{Introduction and outline}
\label{Introduction}

Studies of the dynamics of $\mathcal N=1$ super Yang-Mills theory (SYM) on $\circle$ has been of much interest in the past couple decades. With supersymmetry preserving boundary conditions along the small spatial cycle of length $L$ there is a smooth interpolation from the effective three dimensional theory and the full four dimensional one, due to the large-$N$/large-volume correspondence\footnote{This correspondence states that for a theory compactified onto an effective lower dimensional theory the expectation values of observables are the same in the higher dimensional non-compactified theory with corrections of the order of $O(1/N^2)$ (as long as centre symmetry is preserved [19], [28]). It is important to note, however, that the regime of interest for the large-$N$/volume correspondence requires $N_c$, or $c_2(G)$, to be large enough that $L\Lambda c_2(G)>>1$, and so to study higher dimensional theories one must forfeit the benefits of semi-classical calculability. Yet there may still be information one can get about the higher dimensional theory for 'intermediate' sized $N$ (with $LN\Lambda$ sufficiently less than one) as the expectation values of observables are correct to $O(1/N^2)$ and so errors fall off quicker than $1/N<<1/L\Lambda$. This will not be discussed further in this work, however.}, and other dualities. Furthermore, at small such $L$, the theory becomes weakly coupled and the non-Abelian supersymmetric theory becomes fully Abelianized. This ensures the existence of no light mass charged matter fields to allow the theory to run into strong coupling in the infrared, large-distance theory, at energies less then $1/NL$\footnote{Or $1/c_2(G)L$ for general groups other than $SU(N)$, where $c_2(G)$ is the dual Coxeter number of the gauge group. See Appendix A for more on Lie group and Lie algebra conventions and notations.}. This is the mass of the lightest W boson of the theory with mass acquired as the lightest excited Kaluza-Klein mode winding the spatial cycle $S^1_L$. Our regime of weak coupling then occurs for $NL\Lambda\ll 1$, where $\Lambda$ is the strong coupling scale where the strong coupling $g^2\approx O(1)$. This allows reliable semi-classical monopole-instanton calculations to be performed, and both perturbative and non-perturbative sectors of the theory can be reliably explored analytically. 
\\

Studies of SYM can also apply to finding results of non-SUSY theories as well [1], such as QCD(adj) (i.e. YM with multiple adjoint Weyl fermions). It is found that in both SUSY and non-SUSY theories that the underlying mechanism of confinement and mass-gap generation is due to the proliferation of certain topological 'excitations' in the vacuum of the theory: correlated instanton-anti-instanton events such as the magnetic and neutral bions. The magnetic bions carry zero topological charge and two units of magnetic charge. It is these magnetic bions that lead to mass gap due to Debye screening in the magnetic bion gas. 
\\

The neutral bions are somewhat more interesting and peculiar. These 'molecules' carry zero units of magnetic and topological charge, but two units of scalar charge. their formation and stability is of interest and results from either evoking supersymmetry, using excluded volume arguments, or the use of a certain Bogomolny-Zinn-Justin prescription [4]. 'Resurgence' theory has its role here. It is these neutral bions that allow for the centre-stabilization of the gauge group in the vacuum of the theory, allowing for the confined phase where centre-symmetry is preserved at low temperatures, even though perturbative effective potentials tend to destabilize centre-symmetry. The perturbative effective potential of the theory tends to attract eigenvalues of the Wilson lines and destroys centre symmetry through eigenvalue clumping. This perturbative effective potential will be found in Section 2. The combined dynamics of both these perturbative objects (the W bosons in the theory, which are produced thermally at finite temperature) and the non-perturbative objects determine the occurrence of the deconfinement phase transition as they compete at different temperatures. The centre-stabilizing bions dominating at lower temperatures, and the proliferation of centre-destabilizing W-bosons at higher temperatures compete and allow for a deconfining phase transition at a temperature where these objects are in equal strength and abundance.
\\

The usefulness of having a compact dimension to analytically study gauge theories in a theoretically-controlled manner at weak coupling for any gauge group has successfully been done recently for super Yang-Mills on $\circle$ with a mass for the gaugino. This is called deformed SYM (SYM*) studied in [2] and the deconfinement phase transition has been found at a critical gaugino mass (depending on the theta angle of the theory) to be first order for all gauge groups other than $SU(2)$ where it is second order [1]. It is conjectured that this zero temperature quantum phase transition is continuously related, as a function of gaugino mass $m$, to the thermal deconfinement transition of pure Yang-Mills as $m\rightarrow\infty$ at some critical deconfinement temperature $T_c$. Much supports this conjecture, including lattice studies where qualitative agreement to the zero temperature phase transition has been shown such as the order of the transition, its universality class of centre-symmetry breaking, and dependence on theta-angle [17].\\

This work considers a finite temperature study of $\mathcal N=1$ super Yang-Mills (SYM) by having an additional compact direction along the time direction of size the inverse temperature of the theory $\beta=1/T$.  Having finite temperature breaks the supersymmetry of the theory if we take the (adjoint) fermions in the theory, the gluinos, to have anti-periodic boundary conditions along the thermal circle, and periodic ones for the gluons. This gives us then a full playground in which to study semi-classically a theory in lesser dimensions at finite temperature and with supersymmetry breaking, and eases the study of the deconfinement phase transition.\\

This study of SYM, for the case of $SU(2)$, was recently done in [1] on $\torus$. Here, the role of perturbative and non-perturbative objects in the deconfinement phase transition, where now there is coupling to a scalar field $\phi$, was determined. The resulting dual Coulomb gas of particles involves the W bosons (and superpartners) as well as the neutral and magnetic bions, but with scalar charges attributed to the W's and the neutral bions. These scalar charges attract like charges (unlike how magnetic and electric charges attract opposites) and introduce instabilities to the theory. This gauge theory subjected to lattice studies in [1] both for the so-called double Coulomb gas and its related 'affine' XY-model with symmetry breaking perturbations and fugacities coupled to the scalar field $\phi$. It is this scalar field that breaks the electric-magnetic (Kramers-Vannier) duality enjoyed by the $SU(2)$ theory [15] without supersymmetry on $\mathbb R^3\times S^1_L$. In this paper I generalize these results and dualities for general gauge group $G$ and suggest methods for future study of the phase transitions involved, especially for future lattice Monte Carlo simulations of the Coulomb gas derived here.

\subsection{Outline and summary}
\label{outline}

The purpose of this paper is to generalize the results of [1] for $SU(2)$ $\mathcal N=1$ super Yang-Mills to general gauge group $G$. In [1] the perturbative and non-perturbative contributions to the effective potential were calculated and dualities to a 'double' Coulomb gas of magnetic and neutral bions, as well as W bosons and their wino superpartners, was derived. This was used in lattice studies along with a dual 'affine' XY-model with symmetry breaking perturbations coupled to a scalar field $\phi$. It is the presence of this scalar field that breaks the electric-magnetic duality found in [15] for the non-supersymmetric theory on $\mathbb R^3\times S^1$, and makes the supersymmetric theory here and in [1] harder to study analytically as now the W-boson's fugacity depends on the scalar field $\phi$. Hence we subjected the dual models of the theory to numerical studies and Monte Carlo simulations. A second order phase transition was observed through these simulations and it would be a good task of future research to subject the Coulomb gases and spin models in this paper to lattice studies as well and to observe a first order phase transition for gauge groups other than $SU(2)$, as found in [2], through analytical methods, studying deformed super Yang-Mills (SYM*) with a mass for the gluino.\\

The paper proceeds as follows: in Section 2 I review the perturbative dynamics of Yang-Mills theory on $\mathbb R^2\times S^1_L\times S^1_{\beta}$ and its $\mathcal N=1$ supersymmetric version, beginning with the zero temperature case of $\mathbb R^3\times S^1_L$ in section 2.1 where I set up a notation valid for all gauge groups $G$. In section 2.2, I develop the $T>0$ theory compactified on the torus $\mathbb R^2\times S^1_L\times S^1_{\beta}$ and calculate the perturbative one-loop effective potential.\\

Section 3 is devoted to a review of the non-perturbative sector of the theory valid for all simple Lie algebras $\mathfrak g=Lie(G)$. Monopole solutions and bion structure will be briefly reviewed with emphasis on general gauge group. The zero-temperature dynamics will be discussed followed by the finite temperature dynamics. The total effective potential $V_T=V_{eff.}^{pert.}+V^{non-pert.}$ on $\mathbb R^2\times S^1_L\times S^1_{\beta}$ will be found and consideration of the superpotential will be done.\\

Section 4 contains a derivation of a dual double Coulomb gas to SYM at finite temperature in general gauge group and the partition function will be calculated and constituents of the gas will be studied. This is a main result of this work and one hopes that simulations of this Coulomb gas in the case of general gauge group can be done in the near future as in [1] for $SU(2)$.\\


The Appendix begins with part A reviewing Lie groups and Lie algebras in general, sets up the notation and definitions and explains the necessary concepts. Part B contains a complete derivation of the finite-temperature perturbative effective potential on $\mathbb R^2\times S^1_L\times S^1_{\beta}$ and more details on the derivation of the dual 'universal' Coulomb gas. Part C reviews monopole solutions (BPS and KK) for all gauge groups.

\section{Perturbative dynamics of SYM on $\mathbb R^2\times S^1_L\times S^1_{\beta}$ and effective potential for all gauge groups}
\label{perturbative}

In this section I examine the perturbative dynamics of $\mathcal N=1$ supersymmetric Yang-Mills theory on $\mathbb R^2\times S^1_L\times S^1_{\beta}$ for any semi-simple Lie group $G$. I begin with a review of the theory at zero temperature and move on to new results of the perturbative dynamics at finite temperature $T>0$. The non-perturbative dynamics will be discussed in the next section.

\subsection{$T=0$ dynamics of super Yang-Mills on $\mathbb R^3\times S^1_L$}
\label{zero-T-pert.}

I consider $\mathcal N=1$ supersymmetric Yang-Mills theory with general gauge group $G$ with a single massless adjoint Weyl fermion (the gaugino/gluino). The action on $\mathbb R^2\times S^1_L\times S^1_{\beta}$ is then
\begin{equation}
S=\int_{\mathbb R^2\times S^1_L\times S^1_{\beta}}tr[\frac{1}{2g^2}F^{MN}F_{MN}+\frac{2i}{g^2}\bar\lambda\bar\sigma^MD_M\lambda],
\end{equation}
where $F^{MN}=F^{MNa}T^a$, where $F_{MN}=\partial_MA_N-\partial_NA_M+ig[A_M,A_N]$, is written in the basis of the generators of the Lie group $G$, and similarly $\lambda=\lambda^aT^a$. $D_M$ is the covariant derivative $\partial_M+igA_M$, and $\sigma_M=(i,\vec\tau)$ and $\bar\sigma_M=(-i,\vec\tau)$ where $\vec\tau$ are the Pauli matrices. I write $\vec x\in\mathbb R^2$ for the non-compact (1-2) spatial directions and have $x_0=x_0+\beta$ and $x_3=x_3+L$ for the compact directions. Both the adjoint fermion and the gauge field (gluon) have periodic boundary conditions along the spatial circle $S^1_L$, but along the thermal cycle $S^1_{\beta}$ the former have anti-periodic boundary conditions while the latter have periodic boundary conditions. At zero temperature $\beta\rightarrow\infty$ and the differing boundary conditions do not matter; hence supersymmetry remains unbroken at zero temperature. Recall that I take $L\Lambda_{QCD}<<1$ so one-loop calculations can be easily done to integrate out Kaluza-Klein modes along the $S^1_L$, which calculates the Coleman-Weinberg (or Gross-Pisarski-Yaffe, GPY [39]) effective potential $V_{eff}^{pert.}(\Omega_L)$. Supersymmetry sets this potential to zero as the determinants of gluon and gluino cancel to all orders in perturbation theory. However, non-perturbative corrections in monopole backgrounds contribute a $V_{eff}^{non pert.}$ and will be found in the next Section.\\

At zero temperature we consider the action (2.1) in the vacuum $<A_3>=<A_3^a>H^a$, where we are in the Cartan subalgebra and the theory is Abelian. In general theories the group $G$ gauge theory is broken by the Higgs field $\langle A_3\rangle$ spontaneously to some smaller subgroup $G\rightarrow H\times U(1)^{r-m}$ where $H$ is a subgroup of rank $m<r$. For example, $SU(N)$ breaks down to $U(1)^{N-1}$ and becomes fully Abelianized.
At zero temperature SUSY ($n_f=1$) we always have full Abelianization and $\Lambda c_2(G)L<<1$. Examining the non-perturbative effective potential later we see that $A_3$'s minima are located at the $rank(G)+1=|Z(G)|$ roots of unity ($<A_3>=<A_3^a>H^a$, where $H^a$ are the Cartan generators of the Lie group $G$) and the $Z(G)^{(L)}$ symmetry is preserved at $T=0$. The action (2.1) then becomes:
\begin{equation}
S_{\beta\rightarrow\infty}=\int_{\mathbb R^2\times S^1_{\beta}}\frac{L}{g^2}tr[-\frac{1}{2}F^{\mu\nu}F_{\mu\nu}+(D_{\mu}A_3)^2+2i\bar\lambda(\bar\sigma^{\mu}D_{\mu}\lambda-i\bar\sigma_3[A_3,\lambda])]
\end{equation}
$$=\frac{L}{g^2}\int_{\mathbb R^2\times S^1_{\beta}}(F^a_{\mu\nu}H^a)^2/4+(\partial_{\mu}A_3^aH^a)^2/2+i\bar\lambda^a\bar\sigma^{\mu}\partial_{\mu}\lambda^a,
$$
where in the second line all fields are in the Cartan subalgebra, so $A_3^aT^a\rightarrow A_3^aH^a$, and similarly for $\vec F_{\mu\nu}$ and $\vec\lambda$ (all massive gauge fields have been integrated out in (2.2) above, leaving only massless fields $\{X\}$ in the path integral $\int\mathcal D\{X\}e^{-S[\{X\}]}$).\\

Let me consider the matter present in this theory. The $r$ components of the gauge field $A_3$ along the Cartan subalgebra direction are massless perturbatively, but acquire a small mass $\approx e^{-8\pi^2/g^2}$ non-perturbatively which will be shown in the next section. The remaining components acquire mass $m_W=\pi/c_2(G)L$, where $c_2(G)$ is the dual Coxeter number of the Lie group. (See Table 4 in Appendix A for Lie algebra conventions and data). These are the W-bosons of the theory. Similarly, the components of the gauginos that don't commute with $<A_3>$ acquire the same mass $m_W$ and these constitute the superpartners of the W-bosons, the winos. We also have the remaining components $A_{\mu}^a$ of the gauge field, or, equivalently, $F_{\mu\nu}^a$. Let us assemble these components into a more compact notation. Define first $r$ fields $\sigma^a$ and combine them into a vector $\vec\sigma$ (from now on the vector notation $\vec v$ denotes $r$-dimensional vectors in the Cartan subalgebra), through Abelian duality:
\begin{equation}
\epsilon_{\mu\nu\lambda}\partial_{\lambda}\vec\sigma\equiv\frac{4\pi L}{g^2}\vec F_{\mu\nu}.
\end{equation}
The components of $\vec \sigma$ form $r$ spin-zero dual photon fields. We also define a scalar field $\vec\phi$ as 
\begin{equation}
\vec\phi\equiv \frac{4\pi L}{g^2}\vec A_3-\frac{4\pi^2}{g^2}\vec \rho,
\end{equation}
so that $\vec\phi=0$ corresponds to $Z(G)^{(L)}$ center symmetry unbroken. Here, $\vec\rho$ is the Weyl vector of the Lie algebra, $\vec\rho=\sum_{i=1}^r\vec\omega_i$, where $\vec\omega_i$ are the fundamental weights\footnote{Do not confuse these with the weights of the fundamental representation. See Appendix A for more on weights.} defined by $\vec\alpha_i\cdot\vec\omega_j=\delta_{ij}$. It turns out that this can be written as a sum of positive roots, $\vec\rho=\frac{1}{2}\sum_{\vec\alpha\in\Delta^+_r}\vec\alpha$. The bosonic part of the Lagrangian from (2.2) can be written compactly as $\mathcal L_{free\ bosonic,\beta\rightarrow\infty}=\frac{g^2}{2(4\pi)^2L}[(\partial_{\mu}\vec\sigma)^2+(\partial_{\mu}\vec\phi)^2]$.\footnote{Note that in supersymmetry we can obtain the kinetic terms for the gluon and gluino from the K\:ahler potential $K=\frac{g^2}{2(4\pi)^2L}\bf{B^{\dagger}B}$, where $\bf{B}$ is a dimensionless chiral superfield with lowest component $\vec\phi-i\vec\sigma$.}\\

One further comment before proceeding is the computation of traces in different representations. In (2.2) I took the gauge fields to lie within the Cartan subalgebra of $\mathfrak g$. As outlined in Appendix A, the weights of representation $\mathcal R$ are eigenvalue sets of the Cartan matrices $H^a$ in representation $\mathcal R$. Hence we have for any field $X$, $tr_{\mathcal R}X=Tr(X^aH^a)=\sum_{\vec w\in\Delta_w}^{\mathcal R}\vec X\cdot\vec w$, where $\Delta_w^{\mathcal R}$ is the set of weights of $\mathcal R$, which for the adjoint representation, as seen in Appendix A, are the set of all roots of $\mathfrak g$. We use this from now on in computing traces. I now turn to the finite temperature dynamics.

\subsection{Finite temperature dynamics of SYM on $\mathbb R^2\times S^1_L\times S^1_{\beta}$ and the effective potential}
\label{GPY}

At finite temperature supersymmetry no longer holds as the anti-periodic boundary conditions for the gluino along the thermal cycle break supersymmetry, and the perturbative effective potential no longer vanishes due to non-cancelling of gauge and gaugino determinants. Hence, at finite temperature, our effective Lagrangian takes the form (in the adjoint representation from now on)
\begin{equation}
\mathcal L=L\int d^2x\sum_{\vec w\in\Delta_w^{adj}}[ F^{ij}F_{ij}/4g^2+ |D_i\vec A_3\cdot\vec w|^2/2g^2+|D_i\vec A_0\cdot\vec w|^2/2g^2+ V_{eff}^{GPY\ \mathbb T^2}(\vec A_0,\vec A_3)+
\end{equation}
$$+V_{non\ pert.}(\vec A_0,\vec A_3)+(fermions)],$$
where the non-perturbative potential $V_{non\ pert.}$ will be found in the next section. In this section I compute the perturbative effective potential of the W-bosons and winos present from the Higgsing due to the compactification of the 4D theory.
\\

On calculating the effective one-loop perturbative potential we integrate out the heavy Kaluza-Klein modes along both directions of the torus $\mathbb T^2$, which is related to computing the determinant of the operator $\mathcal O=D_M^2$ on $S^1_L\times S^1_{\beta}$ for both gauge and gaugino fluctuations in the background of constant holonomies along the torus to give us $V_{eff}^{pert}(A_0^a,A_3^a)$. The other components of the gauge field not along the Cartan subalgebra are set to zero in the effective potential as it is minimized by commuting holonomies. I present a method of calculation using zeta functions in Appendix B.1.\\

The end result, combining contributions from both the gauge fields and the gauginos, is
\begin{equation}
V_{eff}^{pert.}=-2\sum_{p=1}^{\infty}[1-(-1)^p]\sum_{\vec w\in\Delta_w}\sum_{n\in\mathbb Z}\frac{e^{-2\pi p|n\beta/L+\beta\vec A_3\cdot\vec w/2\pi|}}{\pi\beta^3Lp^3}(1+2\pi p|n\beta/L+\beta\vec A_3\cdot\vec w/2\pi|)\cos(p\beta\vec A_0\cdot\vec w).
\end{equation}

I now look at the low temperature contribution and consider just the $p=1$ term as other terms are suppressed by higher powers of the Boltzmann factor $e^{-m_W/T}=e^{-\beta/L}$. The result is
\begin{equation}
V_{eff}^{pert., low T}(\vec A_0,\vec A_3)\approx -4\sum_{\vec w\in \Delta_w}\sum_{n\in\mathbb Z}\frac{e^{-2\pi |n\beta/L+\beta\vec A_3\cdot\vec w/2\pi|}}{\pi\beta^3L}(1+2\pi |n\beta/L+\beta\vec A_3\cdot\vec w/2\pi|)\cos(\beta\vec A_0\cdot\vec w).
\end{equation}
Let me make a few comments about the perturbative effective potential. The effective potential is periodic in $\vec A_0$ and $\vec A_3$, with respective periods $2\pi/\beta$ and $2\pi/L$. These fields encode the centre symmetries $Z^{(\beta)}$ and $Z^{(L)}$. Although the $Z^{(\beta)}$ centre symmetry breaks at the deconfinement temperature, the $Z^{(L)}$ centre symmetry remains unbroken up until temperatures beyond the deconfinement phase transition, yet still breaks at a temperature of $T\approx>M_W=2\pi/c_2(G)L$. Also, (2.18) is valid for $c_2(G)LT<<1$ and the mass of the scalar fields $\phi$ (i.e. $\vec A_3$) is exponentially suppressed by the Boltzmann factor $e^{-2\pi/c_2(G)LT}$. At $T=0$ the scalar field is not massless, however, as we will find later the field gets an exponentially small mass from non-perturbative contributions $\approx e^{-4\pi^2/g^2}$, and these dominate at very low temperature. Also note that finding the vacuum of the theory is not found by simply minimizing the perturbative effective potential (2.18) alone. When the non-perturbative sector of the theory, discussed next, is taken into account the thermal electric charges couple to magnetic charges and the total effective potential gets a non-perturbative contribution.

\section{Review of non-perturbative dynamics of SYM on $\mathbb R^2\times S^1_L\times S^1_{\beta}$ for general gauge group }
\label{non-perturbative}

I now examine the non-perturbative sector of the theory. This includes the effects of magnetic monopole-instantons ((anti) self-dual objects) which are charged in the co-root lattice $\Lambda_r^{\vee}$ of the Lie algebra $\mathfrak g$, as well as exotic topological 'molecules': the neutral bions and the the magnetic bions (non self-dual objects). These enter into the path integral with action (3), and hence into the partition function of the theory. I begin by describing the zero temperature dynamics of such particles.\\

Due to the topology of gauge groups we find that for a gauge group $G$ fully abelianized to $U(1)^r$ there are $r$ BPS monopole solutions as $\pi_2(G/U(1)^r)\approx\pi_1(U(1)^r)\approx\mathbb Z^r$. Also, due to the compactness of the $x^3$-coordinate we obtain another solution called the twisted or KK monopole. These solutions are (anti) self-dual objects localized in space and time, and are dilute so we can ignore their internal structure and examine their long-range fields. The field of a single BPS ($\bar{BPS}$) monopole of type $j$, $j=1,\ldots,r$, localized at the origin is, in the stringy gauge, [2] (Here $A_{\mu}=A_{\mu}^aH^a$)
\begin{equation}
A_0^{j,BPS,\bar{BPS}}=\mp \frac{x_1}{r(r+x_2)}\vec\alpha_j^{\vee}\cdot\vec H,
\end{equation}
$$A_1^{j,BPS,\bar{BPS}}=\pm \frac{x_0}{r(r+x_2)}\vec\alpha_j^{\vee}\cdot\vec H,$$
$$A_2^{j,BPS,\bar{BPS}}=0,$$
$$A_3^{j,BPS,\bar{BPS}}=(\frac{\pi}{L}-\frac{1}{r})\vec\alpha_j^{\vee}\cdot \vec H,$$
where $r=\sqrt{x_0^2+x_1^2+x_2^2}$. These gauge field components are charged under the co-root lattice of the Lie algebra and so are multiplied by $\vec\alpha_a^{\vee}$\footnote{The co-roots in other literature are sometimes written as $\vec\alpha_j^*$}. These components of the gauge field give the correct asymptotics of the magnetic field at infinity as in the Appendix (C.11).
The field for the KK ($\bar{KK}$) monopole similarly reads
\begin{equation}
A_0^{0,KK,\bar{KK}}=\pm \frac{x_1}{r(r+x_2)}\vec\alpha_0^{\vee}\cdot\vec H,
\end{equation}
$$A_1^{0,KK,\bar{KK}}=\mp \frac{x_0}{r(r+x_2)}\vec\alpha_0^{\vee}\cdot\vec H,$$
$$A_2^{0,KK,\bar{KK}}=0,$$
$$A_3^{0,KK,\bar{KK}}=(\frac{\pi}{L}+\frac{1}{r})\vec\alpha_0^{\vee}\cdot \vec H.$$

These gauge fields have an additional charge factor $\vec\alpha_0^{\vee}$, the affine co-root of $\mathfrak g$. See Appendix C for more on monopole-instanton solutions. These monopole-instantons carry magnetic charge $Q_m^a$ from Gauss' law
\begin{equation}
\int_{S^2_{\infty}}d^2\Sigma_{\mu}B_{\mu}^j=4\pi Q_m^j,\ j=0,\ldots, r,
\end{equation}
where $B_{\mu}^j=\epsilon_{\mu\nu\lambda}\partial_{\nu}A_{\lambda}^j=Q_m^j\frac{x_{\mu}}{r^3}$ is the magnetic field. I will write $Q_m^i=q_m^i\vec\alpha_i^{\vee}$ as the monopole charges belong to the co-root lattice of the Lie algebra of the gauge group, $\Gamma_r^{\vee}$. $q_m^i$ is the charge of the monopole and equals $\pm 1$. Monopoles of charge type $a$ and $b$ only interact when $\vec\alpha_a^{\vee}\cdot\vec\alpha_b^{\vee}\neq 0$, or, in other words, $a=b$ or they are nearby neighbours on the Dynkin diagram of $\mathfrak g$ (i.e. non-zero elements of the Cartan matrix of the Lie algebra). See Appendix A for the Dynkin diagrams for each (affine) Lie algebra. As mentioned in the introduction, there is also a long-range scalar field (from the $A_3^a$ component of the gauge field) which can attract or repel these monopoles due to scalar charge interaction. There is further a topological charge of these monopole instantons $Q_T$ defined by
\begin{equation}
Q_T^{(i)}=(32\pi)^{-1}\int_{\mathbb R^3\times S^1}F^a_{MN}F^{aMN},.
\end{equation}
Using the solutions (3.1) and (3.2) I find the charges $(Q_m,Q_T)$ for each monopole type, which for $SU(2)$ are:
\begin{equation}
BPS\ (+1,1/2)\ \ \bar{BPS}\ (-1,-1/2)\ \ KK\ (-1,1/2)\ \ \bar{KK}\ (+1,-1/2).
\end{equation}
Note that in general, as is seen in [2], the values of the topological charge depend on the vacuum of the theory. There it is derived that the topological and magnetic charges, for monopoles of type $i$, are
$$Q_T^{(i)}=-\frac{L}{2\pi}\sum_{w\in\Delta_w^{adj}}(\vec w\cdot\vec\phi_0)(\vec w\cdot \vec\alpha_i^{\vee}),\ \ \ \ \ \vec Q_m^{(i)}=\vec\alpha_i^{\vee}.
$$ The topological charge clearly depends on the vacuum $\vec\phi_0$ of the theory, and usually gives fractional charges.
\\

Due to the presence of fermions and supersymmetry (our gaugino), the Callias index theorem [8], [10] on $\mathbb R^3\times S^1$ implies the existence of two adjoint fermionic zero modes attached to each monopole-instanton. Let us use the fields $\vec \phi$ instead of $\vec A_3^a$ and $\vec\sigma$ instead of $\vec A_{\mu}$ and attach fermionic zero modes to get the 't Hooft vertices (the field $\vec z=\vec\phi+i\vec \sigma$ is the lowest component of the chiral superfield $\vec X$ [2])
\begin{equation}
\mathcal M_{BPS,j}=e^{-4\pi^2/g^2}e^{-\vec\alpha_j^{\vee}\cdot\vec z}\vec\alpha_j^{\vee}\cdot\bar{\vec\lambda}\vec\alpha_j^{\vee}\cdot\bar{\vec\lambda},\ \ \ \mathcal M_{KK}=e^{-4\pi^2/g^2}e^{-\vec\alpha_0^{\vee}\cdot\vec z}\vec\alpha_0^{\vee}\cdot\bar{\vec\lambda}\vec\alpha_0^{\vee}\cdot\bar{\vec\lambda}.
\end{equation}
The anti monopole vertices are just the complex conjugates of these. These are valid for arbitrary gauge group due to the supersymmetry of the theory, and are derived in detail from the superpotential in [2]. Inserting these into the partition function of the theory inserts the contribution of fermionic zero modes and the long range fields (the $e^{-\phi^a+i\sigma^a}$ factors). These in themselves due not alter the vacuum structure of the theory as they are attached to fermionic zero modes and do not generate a potential for the fields $\vec \phi$ and $\vec\sigma$ and no mass will be generated for the dual photons $\vec\sigma$. I consider then the effect of the neutral and magnetic bions in the non-perturbative potential and their role in the deconfinement phase transition.\\

Let me now consider the non self-dual 'molecules' these monopole constituents can form. Their charges and amplitudes of the so-called magnetic and neutral bions that form are summarized below in Table 2. In all cases the index $a\in\{0,\ldots,r\}$. A neutral KK bion would have $a=0$ and is formed from a KK-anti-KK monopole pair. In the case of magnetic bions they are formed from two (BPS or KK) monopoles of charge types $a\neq b$ as long as $\vec\alpha_a\cdot\vec\alpha_b\neq 0$, that is the monopoles are Dynkin neighbours on the Dynkin diagram of $G$.\\

\begin{tabular}{l*{6}{c}r}
\rm{Molecule} & \rm{vertex} & $(Q_m.Q_T)$ & \rm{amplitude}\\
\hline
\rm{neutral bion} & $\mathcal M_{a}\bar{\mathcal M}_{a}$ & $(0,0)$ & $e^{-8\pi^2/g^2}e^{-2\vec\alpha_a^{\vee}\cdot\vec\phi}$\\
\rm{magnetic bion} & $\mathcal M_{a}{\mathcal M}_{b}$ & $(2,0)$ & $e^{-8\pi^2/g^2}e^{-(\vec\alpha_a^{\vee}+\vec\alpha_b^{\vee})\cdot\vec\phi-i(\vec\alpha_a^{\vee}-\vec\alpha_b^{\vee})\cdot\vec\sigma}$\\
\hline
\end{tabular}
\\

{\bf Table 2:} Magnetic molecule vertices, charges, and amplitudes for different molecules.\\


Note that these bion factors are for monopole constituents of charge type $a$ and $b$ (or 0) if they can form such bion molecules, that is the charges are the same type, or are Dynkin neighbours. Note for the above cases we can have $a$ or $b=0$ allowing for molecules containing KK monopoles. The occurences (if they occur) of $\phi^0$, $\sigma^0$ can be written in terms the $\phi^a$, $\sigma^a$ as the linear combination $\sum_{a=0}^rk_a^{\vee}\vec\alpha_a^{\vee}=0$, with $k_0^{\vee}\equiv 1$, requires that $\phi^0=-\sum_{a=1}^rk_a\phi^a$, where $k_a^{\vee}$ are the dual Kac labels of the roots of the Lie algebra $\mathfrak g$. The anti-bions are just the complex conjugates of these amplitudes, and have the negative of the charges of the bion. Figure 1 shows an example magnetic bion in the case of $SU(2)$. Since the magnetic bion carries no fermionic zero modes, it generates a potential for the fields $\vec\phi$ and $\vec\sigma$ and gives a mass to the dual photon fields $\vec\sigma$ and the theory can confine electric charges. These magnetic bions are stabilized by the attractive force due to exchange of adjoint fermionic zero modes giving them an effective radius $r_*=4\pi L/g^2$. See [22] for more on bion structure. The fields of the magnetic bions can be found simply by adding the fields (3.1) and (3.2). For example, for the magnetic bion,
\begin{equation}
A_0^{ab,bion}=-2\frac{x_1}{r(r+x_2)}(\vec\alpha_a^{\vee}-\vec\alpha_b^{\vee})\cdot\vec H,
\end{equation}
$$A_1^{ab,bion}=2\frac{x_0}{r(r+x_2)}(\vec\alpha_a^{\vee}-\vec\alpha_b^{\vee})\cdot\vec H,$$
$$A_2^{ab,bion}=0,$$
$$A_3^{ab,bion}=\frac{2\pi}{L}(\vec\alpha_a^{\vee}-\vec\alpha_b^{\vee})\cdot\vec H.$$
These gauge fields are accompanied by the differences of charges $\vec\alpha_a^{\vee}-\vec\alpha_b^{\vee}$. Neutral bions involve combinations of monopoles and antimonopoles of the same charge type, but have scalar rather than magnetic charges. One would then replace the occurences of $(\vec\alpha_a^{\vee}-\vec\alpha_b^{\vee})$ with the imaginary charge $2i\vec\alpha_i^{\vee}$ in the imaginary co-root lattice of $G$, $i\Lambda_r^{\vee}$, for the neutral bion fields. This imaginary charge is what makes like scalar charged neutral bions attract as opposed to likely charged magnetic bions which repell.
\\

The neutral bions, however, are a little more tricky to see how they form. Usually an analytic continuation is required (a so-called BZJ prescription) to control the attractive forces of the monopole constituents, or a finite volume argument to make sure they are stable with finite size [4], [22]. Supersymmetry can also be invoked [2] to allow for their stability. 'Resurgence' theory has a role here [4]. Nonetheless, these objects are stable and generate a centre-stabilizing potential for confinement. Since the perturbative potential vanishes, only the neutral bion-induced potential can lead to centre-stabilization.\\

The total effective potential of the non-perturbative contributions $V_{eff}^{non\ pert.}(\vec\phi,\vec\sigma)$ is found by adding the amplitudes in (3.7). As in [2] we can obtain the effective potential from the superpotential in terms of the (chiral) superfield $\vec X$,
\begin{equation}
\mathcal W=\kappa\frac{L}{g^2}\mu^3(\sum_{j=1}^r\frac{2}{\vec\alpha_j^2}e^{\vec\alpha_j^{\vee}\cdot\vec X}+\frac{2}{\vec\alpha_0^2}e^{\vec\alpha_0^{\vee}\cdot\vec X+2\pi i\tau}),
\end{equation}
where $\kappa$ is a numerical factor that will not matter to us, $\vec X$ is the chiral superfield with lowest component $\vec\phi-i\vec\sigma$ and $\tau\equiv i\frac{4\pi^2}{g^2}+\theta/2\pi$ will be taken with $\theta=0$ here. In [2] it is shown that quantum corrections to the superpotential change the scale of the coupling to be not $\Lambda_{PV}$ but rather $\mu=2/R$, with $L=2\pi R$. The effective action is then found from (see [31] for more on superpotentials and supersymmetry)
\begin{equation}
S(\vec\phi,\vec\sigma)=\int d^3x[K_{i\bar{j}}\partial_{\mu}X^i\partial^{\mu}X^{\dagger \bar{j}}+K^{i\bar{j}}\frac{\partial\mathcal W}{\partial X^i}\frac{\partial\bar{\mathcal W}}{\partial X^{\dagger j}}],
\end{equation}
where the K\:ahler potential is found from [2], to one loop quantum corrections, $K^{i\bar{j}}=\frac{16\pi^2L}{g^2}[\delta_{ij}-\frac{3g^2}{16\pi^2}\sum_{\vec w\in\Delta_{w+}^{\mathcal R}}w_iw_j[\psi(\vec w\cdot\vec\phi/2\pi)+\psi(1-\vec w\cdot\vec\phi/2\pi)]]$, where $\psi(z)=\Gamma'(z)/\Gamma(z)$ is the digamma function. The inverse K\:ahler metric has the overall coefficient inverted and the second term becomes negative. We shall ignore the one-loop corrections for now in this paper. The non-perturbative effective potential due to bions is from the second term of (3.9) and so we get (ignoring the quantum one-loop corrections to the K\:ahler potential in the sequel)
\begin{equation}
V_{bion}=\frac{16\pi^2L}{g^2}\delta^{ij}\frac{\partial\mathcal W}{\partial X^i}\frac{\partial\bar{\mathcal W}}{\partial X^{\dagger j}}=64\pi^2\kappa^2(\frac{2\pi R}{g^2})^3(\frac{2}{R})^6[\sum_{i,j=1}^r\frac{\vec\alpha_i^{\vee}\cdot\vec\alpha_j^{\vee}}{\vec\alpha_i^2\vec\alpha_j^2}e^{\vec\alpha_i^{\vee}\cdot\vec X+\vec\alpha_j^{\vee}\vec X^{\dagger}}+
\end{equation}
$$\frac{\vec\alpha_0^{\vee 2}}{\vec\alpha_0^4}e^{\vec\alpha_0^{\vee}\cdot(\vec X+\vec X^{\dagger})+2\pi i(\tau-\tau^*)}+\sum_{i=1}^r\frac{\vec\alpha_i^{\vee}\cdot\vec\alpha_0^{\vee}}{\vec\alpha_i^2\vec\alpha_0^2}(e^{\vec\alpha_i^{\vee}\cdot\vec X+\vec\alpha_0^{\vee}\vec X^{\dagger}-2\pi i\tau^*}+e^{\vec\alpha_0^{\vee}\cdot\vec X+\vec\alpha_i^{\vee}\vec X^{\dagger}+2\pi i\tau})],
$$
where $\vec X=i(\tau\vec\phi+\vec\sigma)-\frac{3}{2}\sum_{\vec w\in\Delta_w^{adj+}}(\vec w\log\frac{\Gamma(\vec w\cdot\vec\phi/2\pi)}{\Gamma(1-\vec w\cdot\vec\phi/2\pi)})$ is the one-loop correction to the superfield derived in [2]. I shall ignore the quantum corrections for now. After some algebra the non-perturbative effective potential becomes
\begin{equation}
V_{bion}=V^0_{bion}\sum_{i,j=0}^rk_i^{\vee}k_j^{\vee}\vec\alpha_i^{\vee}\cdot\vec\alpha_j^{\vee}e^{-(\vec\alpha_i^{\vee}+\vec\alpha_j^{\vee})\cdot\vec b}\cos(\vec\alpha_i^{\vee}-\vec\alpha_j^{\vee})\cdot\vec\sigma',
\end{equation}
where $\vec\sigma'=\vec\sigma-\vec\sigma_0$ and $\vec b=\frac{4\pi}{g^2}(\vec\phi-\vec\phi_0)$ are the fluctuations about the supersymmetric vacuum, $\vec\phi_0,\vec\sigma_0$, and
\begin{equation}
V^0_{bion}=16\pi^2\kappa^2(\frac{512\pi^3}{g^6R^3|v|^2})e^{-16\pi^2/g^2c_2(G)},
\end{equation}
with $|v|=[\prod_{i=0}^r(\frac{k_i^{\vee}\vec\alpha_i^2}{2})^{k_i^{\vee}}]^{1/c_2(G)}$.
Note that this gives the $SU(2)$ result with $\alpha_0=-\alpha_1$ leading to $\cosh 2\phi-\cos 2\sigma$ terms as found in the non-perturbative effective potential in [1].

The monopole terms carry two fermionic zero modes (the $\lambda$'s in (3.6)), and will not be considered here as there is no mass gap generation from such terms. Hence we consider the W-boson/photon and magnetic and neutral bion contributions to the effective potentials in (2.5).\\

The idea at zero temperature is that the $Z(G)_L$ centre symmetry is unbroken 
as $<\vec\phi>=0$ minimizes the potential in (3.12) (at zero temperature, and low T in general, we can ignore the perturbative, deconfining, effective potential (2.18) due to Boltzmann supression of the W-bosons). There is also a mass gap for the dual photon, the $\phi^a$'s and $\lambda^a$'s have equal masses and the electric charges are confined.


\subsection{Monopole-instantons and bion structure at finite temperature}
\label{finite T monopole}

From [1] let me recall that the dual photon and $\phi$ fields have masses $m_{\sigma^a}=m_{\phi^a}\approx e^{-4\pi^2/g^2}/L$, and the fermions acquire a thermal mass $\approx T$. The light gauginos do not participate in the deconfinement phase transition as they carry no electric, magnetic or scalar charge, however they do allow the formation of the bions and so have an indirect role in the transition. The heavy fermions (the winos) do participate, however, in a way similar to the W-bosons. The deconfinement transition temperature is found to be of the order $T_c\approx g^2/8\pi L$ and $~\Lambda_{QCD}$ from both simulations and from setting the W-boson and magnetic bion fugacities to the same order [15]. This temperature is smaller then the inverse bion radius and so we need not worry about the dissolution of the bions before the deconfinement transition and our gas of such particles exists beyond the deconfinement transition temperature. See [1] for more details.\\
The best way to study the finite temperature dynamics of our gas of all these particles is to map it to a double Coulomb gas and examine its partition function. This I derive in the next section.

\section{Duality to dual double Coulomb gas}
\label{Coulomb gas}

To derive our electric-magnetic Coulomb gas dual to our theory I will follow a derivation done in [6] for the finite temperature 3D Polyakov model. We go back to our field $F_{\mu\nu}^a$ instead of $\sigma^a$ and add to the perturbative photon fluctuations the contribution of the magnetic field of our magnetic bions (and anti magnetic bions). This involves calculating the electric W-boson (and wino) determinant in the multi-instanton/anti-instanton background. We can similarly find the W-boson determinant in the background of the neutral bions (both thw W's and neutral bions carry scalar charge due to their coupling to the scalar fields $\phi^a$). In the end we find a duality of our model to a dual double electric-magnetic Coulomb gas with also scalar charges which couple to the fields.\\

Hence we begin by splitting the fields into photon fluctuation and magnetic field components
\begin{equation}
F^a_{\mu\nu}=F^{a,bion}_{\mu\nu}+F^{a,ph}_{\mu\nu}
\end{equation}
$$A^a_{\mu}=A^{a,bion}_{\mu}+A^{a,ph}_{\mu},$$
where $A^{a,bion}_{\mu}=\sum_{i,q_i=\pm 1}q_iA_{\mu}^{a,bion}(x-x_i)$ splits into a sum of an arbitrary numbers of bions and anti bions at positions $x_i\in\mathbb R^3$, and $A_{\mu}^{a,bion}$ is from (26). At finite temperature we have $\beta$ finite and so we must sum up an infinite number of image charges in the 0-direction and so
\begin{equation}
A^{a,bion}_{\mu}=\sum_{a,q_a=\pm 1}\sum_{n\in\mathbb Z}q_aA_{\mu}^{a,bion}(\vec x-\vec x_a,x_0-x_{0,a}+n\beta).
\end{equation}
The partition function of our Coulomb gas is the path integral of our field theory with path integrals over the gauge fields $A_{\mu}^{a,ph}$ and the scalar fields $\phi^a$, and sums over arbitrary numbers $N_{b\pm}$ of magnetic bions, as well as $N_W$ W-bosons and their superpartners. There is also a sum over 'colours', that is the sum over components of the Cartan subalgebra of $\mathfrak g$. In the action we integrate the photon fields in the background of magnetic (and neutral) bions. Let us first consider the photon field in the background of magnetic bions. We simply replace the argument of the cosine of the potential (2.18) with the integral
\begin{equation}
\oint_{S^1_{\beta}}dx_0A_0^{a,bion}=\sum_{a,q_a=\pm 1}q_i\int_0^{\beta}\sum_{n\in\mathbb Z}A_{\mu}^{a,bion}(\vec x-\vec x_a,x_0-x_{0,a}+n\beta)=\sum_{a,q_a=\pm 1}q_a\int_{-\infty}^{\infty}A^{a,bion}_0(\vec x-\vec x_a,x_0),
\end{equation}
which, using equation (3.8) we find as in [1] the integral above to be
\begin{equation}
\oint_{S^1_{\beta}}dx_0A_0^{bion,ij}=2\sum_{i,j=0}^r\sum_{a,q_a=\pm}(\vec\alpha_i^{\vee}-\vec\alpha_j^{\vee})q_a\Theta(\vec x-\vec x_a),
\end{equation}
where $\Theta(\vec x)=-sgn(x_1)\pi/2+\tan^{-1}\frac{x_2}{x_1}$ is the angle in the $x_1$-$x_2$-plane between $\vec x$ and the magnetic bion of type $ij$ at position $\vec x_a$. From here on $A,B$ denotes the position of the W-bosons, $a,b$ the positions of the magnetic bions.\\

Using this term in the cosine in the action with GPY potential (2.18) I can write the action of the partition function in the grand canonical ensemble for bions at positions $a$ as
\begin{equation}
S=\int_{S^1_{\beta}\times\mathbb R^2\times S^1_{L}}\frac{L}{2g^2}(\partial_{\mu}\vec\phi)^2-\frac{L}{4g^2}(\vec F_{\mu\nu}^{ph}+\vec F_{\mu\nu}^{bion})^2
\end{equation}
$$-\sum_a^{N_{b+}+N_{b-}}2T\xi_W(\vec\phi)\cos(2\sum_{i,j,k}\sum_{a,q_a=\pm}(\vec\alpha_i^{\vee}-\vec\alpha_j^{\vee})\cdot\vec\alpha_kq_a\Theta(\vec x-\vec x_a)+\oint_{S^1_{\beta}}dx_0A_0^{a,ph})+V_{neutral\ bion}(\vec \phi(x_a)),$$
where $V_{neutral\ bion}(\vec x_a)$ is given by (3.11) for like charges $i=j$, with the fields at $\vec x_a$. Note the dependence of the W-boson fugacity, $\xi_W$, on $\vec\phi$. The grand canonical partition function $\mathcal Z_{grand}=\int\mathcal D\vec\phi\int\mathcal D \vec A_{\mu}^{ph}e^{-S}$ can be expanded using the equation
\begin{equation}
\exp[2\xi\int dx\cos(f(x))]=\sum_{n_{\pm}=0}^{\infty}\sum_{q_i=\pm 1}\frac{\xi^{n_{+}+n_{-}}}{(n_{+})!(n_{-})!}\prod_{i=1}^{n_{+}+n_{-}}\int dx_i e^{\sum_iiq_if(x_i)},
\end{equation}
and including the potential of the neutral bions $i=j$, $V_{neutral\ bion}(\vec\phi)$ I rewrite the partition function as
\begin{equation}
\mathcal Z_{grand}=\sum_{N_{b\pm},q_i=\pm}\sum_{N_{W\pm},q_A=\pm}\prod_a^{N_{b+}+N_{b-}}\int d^3x_a\prod_A^{N_{W+}+N_{W-}}\int d^3x_A\int\mathcal D \vec\phi\int\mathcal D A_{\mu}^{a,ph}\frac{\xi_b(\phi^a)^{N_{b+}+N_{b-}}}{N_{b+}!N_{b-}!}\times
\end{equation}
$$\frac{(T\xi_W(\phi^a))^{N_{W+}+N_{W-}}}{N_{W+}!N_{W-}!}\exp[\int_{\mathbb R^3\times S^1_{\beta}}\frac{L}{2g^2}(\partial\vec\phi)^2+2i\sum_{i,j,q_a=\pm}(\vec\alpha_i^{\vee}-\vec\alpha_j^{\vee})q_a\Theta(\vec x_A-\vec x_a)-\int_{\mathbb R^3\times S^1_{\beta}}\frac{L}{4g^2}(\vec F_{\mu\nu}^{ph}+\vec F_{\mu\nu}^{bion})^2$$ $$-i\sum_A\sum_{a=0}^rq_A\vec\alpha_a\cdot\vec A_0^{ph}(\vec x,x_0)\delta(\vec x-\vec x_A)+V_{neutral\ bion}(\vec\phi)].
$$
It is clear that by writing a path integral $\int\mathcal D\vec\phi$ (or $\int\mathcal D A_{\mu}^{a,ph}$ through (2.4)) is equivalent to integrating over $\vec b$ introduced earlier. Also I wrote the bion fugacity as $\xi_b(\phi^a)=V^0_{bion}=16\pi^2\kappa^2(\frac{512\pi^3}{g^6R^3|v|^2})e^{-16\pi^2/g^2c_2(G)}$, and the fugacity of the W-bosons is from the Boltzmann distribution $\xi_W=2\int\frac{d^2p}{(2\pi)^2}e^{-m_W/T-p^2/2m_WT}=2\frac{Tm_W}{2\pi}e^{-m_W/T}$, and $m_W=1/c_2(G)L$ is the mass of the lightest W-bosons. This W-boson fugacity actually depends on $\phi^a$ in general, as found from the calculation (2.18) of the W-boson determinant. We have
\begin{equation}
\xi_W(\phi^a)=\frac{2}{\pi\beta^2}\sum_{n\in\mathbb Z}\sum_{\vec w\in\Delta_w^{\mathcal R}}e^{-\frac{\beta}{L}|(2n+1)\pi+g^2\vec\phi\cdot\vec w/4\pi|}(1+\frac{\beta}{L}|(2n+1)\pi+g^2\vec\phi\cdot\vec w/4\pi|)
\end{equation}
$$=\sum_{\vec w\in\Delta_w^{\mathcal R}}\frac{2}{\beta L\sinh (\beta\pi/L)}[(\coth\frac{\beta\pi}{L}+\frac{L}{\pi\beta})\cosh\frac{\beta g^2\vec\phi\cdot\vec w}{4\pi L}-\frac{g^2\vec\phi\cdot\vec w}{4\pi^2}\sinh\frac{\beta g^2\vec\phi\cdot \vec w}{4\pi L}],
$$
where I assume $\phi^a$ lies in the Weyl chambre $\vec\alpha_i\cdot\vec\phi\geq 0$ for each $\vec\alpha_i$ simple, and $-\vec\alpha_0\cdot\vec\phi\leq 1$. For small values of $\phi^a\approx 0$ near its minimum I find the contribution to $\xi_W(\vec\phi)$ is dominated by the $n=0,-1$ terms and we are left with (at the minimum $\phi^a=0$) $\xi_W\approx\frac{4}{\beta L}e^{-\beta\pi/L}=\frac{4c_2(G)m_WT}{\pi}e^{-c_2(G)m_W/T}$, as expected, but we get $2c_2(G)$ times the usual Boltzmann factor due to the fact that there are $c_2(G)$ lightest Kaluza-Klein modes with the same mass $m_W=\pi/c_2(G)L$ (due to the unbroken centre symmetry), and the wino superpartners have the same mass as well and contribute equally. I also represent the charges as vectors $q_{X,i}=\vec\alpha_i^{(\vee)}$ for electric and magnetic charges, respectively, for charge type ('colour') $i$. The last term in the exponent of (4.7) encompasses all neutral bions introduced in the previous section.\\

Next, a duality transformation can be done as in [6] for the gauge field $A_{\mu}^{a,ph}$ (and consider only the zero mode of its $\beta$-component in the low-$T$ approximation) to exchange the gauge fields $F_{\mu\nu}^a$ for scalar fields $\sigma^a$. This goes as follows:\\

The classical free part of the action can be written as (with theta-term added for generality)
\begin{equation}
S_{cl}=\frac{L}{g^2}\int d^3x[\frac{1}{L^2}(\partial_{\mu}\vec\phi)^2-\frac{1}{2}(\vec F_{\mu\nu})^2+2i\vec\bar\lambda\cdot\bar\sigma_{\mu}D_{\mu}\vec\lambda]-i\frac{\theta}{8\pi^2}\int d^3x\epsilon_{\mu\nu\rho}\partial_{\mu}\vec\phi\cdot\vec F_{\nu\rho}.
\end{equation}
We then add a set of $r$ Lagrange multiplier fields $\sigma^a$ to impose the Bianchi identity on the $r$ components of $\vec F_{\mu\nu}$ and construct the constraint part of the action
$$S_{aux}=\frac{i}{4\pi}\int d^3x\vec\sigma\epsilon_{\mu\nu\rho}\cdot\partial_{\mu}\vec F_{\nu\rho}=-\frac{i}{2\pi}\oint_{S^2_{\infty}}d\Sigma_{\mu}\vec\sigma\cdot\vec B_{\mu},
$$
where the second line is by integration by parts. Integrating out $\vec\sigma$ clearly gives the Bianchi identity $\epsilon_{\mu\nu\rho}\partial_{\mu}\vec F_{\nu\rho}=0$, and integrating out $\vec F_{\mu\nu}$ gives a dual description for the action, whose free bosonic part now reads
\begin{equation}
S_{free-bosonic}=\frac{1}{L}\int d^3x[\frac{1}{g^2}(\partial_{\mu}\vec\phi)^2+\frac{g^2}{16\pi^2}(\partial_{\mu}\vec\sigma+\frac{\theta}{2\pi}\partial_{\mu}\vec\phi)^2]=(4\pi L)^{-1}\int d^3x\partial^{\mu}\vec z^*\cdot\partial_{\mu}\vec z/Im\tau,
\end{equation}
where $\vec z=i(\tau\vec\phi+\vec\sigma)$ is the lowest component of the superfield $\vec X$ with fermionic component $\vec\lambda$. We take $\theta=0$ in the sequel. Using this duality, the free bosonic part of our Lagrangian (2.5) becomes:
\begin{equation}
S_{free-bosonic}=\beta\int d^2x\frac{L}{2g^2}[(\partial_{\mu}\vec\phi)^2+(\partial_{\mu}\vec\sigma)^2].
\end{equation}

The derivation of the Coulomb gas partition function continues with a series of dualities, which are described in Appendix B.2. The end result for the partition function of our dual Coulomb gas is the final result
\begin{equation}
\mathcal Z_{grand}=\sum_{N_{b\pm},q_i^a=\pm}\sum_{N_{W\pm},q_A^a=\pm}\frac{\xi_b^{N_{b+}+N_{b-}}}{N_{b+}!N_{b-}!}\prod_i^{N_{b+}+N_{b-}}\int\prod_{a=0}^r d^3x_i^a\frac{(T\xi_W(\phi))^{N_{W+}+N_{W-}}}{N_{W+}!N_{W-}!}\prod_A^{N_{W+}+N_{W-}}
\end{equation}
$$\int \prod_{a=0}^rd^3x_A^a\int\mathcal D\vec\phi\exp[\sum_{a,b=0}^r(\frac{32\pi LT}{g^2}\sum_{i>j}q_i^aq_j^b\vec\alpha_a^{\vee}\cdot\vec\alpha_b^{\vee}\log|\vec x_i^a-\vec x_j^b|+\frac{g^2}{2\pi LT}\sum_{A>B}q_A^aq_B^b\vec\alpha_a\cdot\vec\alpha_b\log|\vec x_A^a-\vec x_B^b|+$$
$$4i\sum_{i,A}q_i^aq_A^b\vec\alpha_a^{\vee}\cdot\vec\alpha_b\Theta(\vec x_A^b-\vec x_i^a))+\int_{\mathbb R^2}[\half\frac{g^2}{(4\pi)^2L}(\partial_{\mu}\vec\phi)^2+V^0_{bion}\sum_{a=0}^r(k_a^{\vee}\vec\alpha_a^{\vee})^2\exp(-\frac{8\pi}{g^2}\vec\alpha_a^{\vee}\cdot\vec\phi)]],$$
which is valid for all $T$ with $0\leq T<M_W$, and I used $g_3=g/L$ and the long-distance property of the Green's function where it behaves like a logarithm. The derivation is reserved for the Appendix but I will make comments here.\\

Note already the usual Coulomb-Coulomb interactions between electric W-bosons and magnetic bions, as well as the Aharanov-Bohm interaction given by the $\Theta$ term as in [1]. One further point to consider is the dependency of the W-boson fugacity on the fields $\phi^a$. In the case of circle compactification where the $\vec\phi$ field is absent (as in the zero temperature limit) we have a Kramers-Wannier duality $32\pi LT/g^2\rightarrow g^2/2\pi LT$ as in the sine-Gordon model which is the zero-temperature limit of our theory. The magnetic monopoles are not present in the partition function of our Coulomb gas, as these we ignore as they do not contribute to the dynamics of the deconfinement phase transition and the vacuum structure of the theory as they carry fermionic zero modes. These monopoles still interact with a potential similar to the W-bosons: $V_{m-m}= \frac{4\pi}{LTg^2}\sum_{i,j=1}^{N_m}\sum_{q_a,q_b=\pm}\sum_{a,b=0}^rq_i^aq_j^b\vec\alpha_a^{\vee}\cdot\vec\alpha_b^{\vee}\ln|\vec x_i^a-\vec x_j^b|$. Note the hierarchy of scales in the effective 2D Coulomb gas: $r_m\approx L<<r_b\approx L/g^2<<d_{m-m}\approx Le^{2\pi^2/g^2}<<d_{b-b}\approx Le^{4\pi^2/g^2}$ of monopole size, bion size, monopole-monopole separation distance, and bion-bion separation distance. This holds at weak coupling and shows that the vacuum partition function is truly that of an effective 2D dilute Coulomb gas of monopoles and bions. Note that the hierarchy fails at strong coupling and the Coulomb gas 'collapses', showing the importance of weak coupling to our duality.\\



The $r$ W-bosons of the theory can be written in terms of the scalar and photon fields:
$$\vec W^{\pm}=\vec\phi\pm i\vec\sigma.$$ These particles can be thought of as having two charges: scalar and electric. The electric charges belong as usual to the root lattice $\Z[\{\vec\alpha_i\}]=\Lambda_r$, whereas the scalar charges belong to the imaginary root lattice $i\Z[\{\vec\alpha_i\}]=i\Lambda_r$ (the magnetic charges belong dually to the co-root lattice of $G$, $\Lambda_r^{\vee}$).\\

To explain this one could go one step further and evaluate the path integral of the $\vec\phi$ field, but this proves difficult for general group. However I will do it for the case of $SU(2)$ in [1], where it was not done. The $SU(2)$ result was in [1]
\begin{equation}
\mathcal Z_{grand}=\sum_{N_{b\pm},q_i=\pm}\sum_{N_{W\pm},q_A=\pm}\frac{\beta\xi_b^{N_{b+}+N_{b-}}}{N_{b+}!N_{b-}!}\prod_i^{N_{b+}+N_{b-}}\int d^2x_i\frac{(\xi_W(\phi))^{N_{W+}+N_{W-}}}{N_{W+}!N_{W-}!}\prod_A^{N_{W+}+N_{W-}}\int d^2x_A\int\mathcal D\phi
\end{equation}
$$\exp[\frac{32\pi LT}{g^2}\sum_{a>b}q_aq_b\log|\vec x_a-\vec x_b|+\frac{g^2}{2\pi LT}\sum_{A>B}q_Aq_B\log|\vec x_A-\vec x_B|+4i\sum_{a,A}q_aq_A\Theta(\vec x_a-\vec x_A)+$$
$$+\int_{\mathbb R^2}d^2x(\half\frac{g^2}{32\pi^2LT}(\partial_{\mu}\phi)^2+\frac{64\pi^2e^{-8\pi^2/g^2}}{TL^3g^6}\cosh 2\phi)].
$$
One can go a step further and expand the cosh term using a result similar to (4.6), and solve the equations of motion for the $\phi$ field in the background of neutral bions. I present the details in Appendix B.2. The result for $SU(2)$ is
\begin{equation}
\mathcal Z_{grand}=\frac{Z_0}{L^2\beta}\sum_{N_b}\sum_{N_W}\sum_{N_{b'}}\sum_{q_X=\pm}\frac{\xi_b^{N_{b+}+N_{b-}}(2\xi_W)^{N_{W+}+N_{W-}}\xi_{b'}^{N_{b'+}+N_{b'-}}}{N_b^+!N_b^-!N_W^+!^2N_W^-!^2N_{b'+}!N_{b'+}!}\times
\end{equation}
$$\times\prod_a^{N_b^++N_b^-}\prod_A^{N_W^++N_W^-}\prod_{\alpha}^{N_{b'}^++N_{b'}^-}\int d^{(2+1)}r_a\int d^{(2+1)}r_A\int d^{(2+1)}r_{\alpha}\times$$
$$\exp[-\frac{1}{\pi g^2}\sum_{a\neq b}q_a^mq_b^m\ln|r_a-r_b|-\frac{g^2}{4\pi(T)}\sum_{A\neq B}(q_A^eq_B^e-q_A^sq_B^s)\ln|r_A-r_B|-4i\sum_{a,A}q_a^mq_A^e\Theta(r_a-r_A)+$$ $$+\frac{16\pi}{g^2}\sum_{\alpha\neq \beta}q_{\alpha}^sq_{\beta}^s\ln|r_{\alpha}-r_{\beta}|+4\sum_{\alpha,A}q_{\alpha}^sq_A^s\ln|r_{\alpha}-r_A|].$$

From the interaction terms in the partition function (4.18) we see that like scalar charges attract, whereas like electric and magnetic charges repel. This is due to the different sign in their interaction. Similarly one could work out the propagator for the sinh-Gordon model and find that it is the same as the sine-Gordon model but with opposite sign. The W-bosons have a double nature attracting opposite electrically charged W-bosons, and attracting like scalar charged W-bosons and neutral bions. This Coulomb gas can be thought of as a 'pansexual-like' gas from these different interactions present. Let us summarize the components in the Coulomb gas and their charges, written as vectors $q_X=(q_{X,e},q_{X,m},q_{X,s})$. For $SU(2)$ there is only one root and co-root and so the charges are written simply as $\pm 1$. For other gauge group a magnetic charge of 2 for magnetic bions corresponds to negative combinations of two Dynkin-neighbouring charges $\vec\alpha_i^{\vee}-\vec\alpha_j^{\vee}$. The neutral bions can be interpreted as having imaginary charge $2i\vec\alpha_i^{\vee}$.\\

\begin{tabular}{l*{6}{c}r}
\rm{Coulomb  gas constituent}  & $q_X=(q_{X,e},q_{X,m},q_{X,s})$ \\
\hline
\rm{magnetic bions} & $(0,\pm 2,0)$\\
\rm{W-bosons} & $(\pm 1,0,\pm 1)$\\
\rm{neutral bions} & $(0,0,\pm 2)$\\
\hline

\end{tabular}
\\

{\bf Table 3:} Scalar, electric and magnetic charges of relevant Coulomb gas constituents.\\

This Coulomb gas can be subjected to lattice study as in [1] for the case of $SU(2)$, but for other gauge groups. Perhaps extending first the results to $SU(3)$ and $SU(N)$ would be a start in future research. See [1], [53], [54] for more on the Monte-Carlo simulations used in studying such Coulomb gases numerically. Another method of studying the deconfinement phase transition other than simulating the Coulomb gas is to map the Coulomb gas constituents to parameters of a dual spin model. The spin model that best suits the Coulomb gas at hand is a multiple component XY spin model with symmetry breaking perturbations and fugacities coupled to the scalar field $\vec\phi$. This is a project for future research as there are difficulties for groups other than $SU(2)$ and $SU(3)$ due to the dependence of both the W-boson and bion fugacities on the scalar field $\vec\phi$. I will not make further comment on spin models in this paper but will point out that the dual Coulomb gas can be simulated just as described in [1] and I hope this can be done in a soon future work.

\section{Conclusions and future work}
\label{conclusion}

It was found that $\mathcal N=1$ super Yang-Mills on $\mathbb R^3\times S^1$ has a dual description as a double Coulomb gas of various particles: W-bosons and their wino superpartners, monopole-instantons and neutral and magnetic bions and their anti-particles. The partition function was computed as well as the duality maps to the Coulomb gas of $r$ such types of electric and magnetic charges, and several types of magnetic and neutral bions formed from combinations of BPS and KK monopoles (and their anti-monopoles). The electric charges are charged under the root lattice of the gauge group $G$, $\Lambda_r$, and the magnetic charges are charged under the co-root lattice, $\Lambda_r^{\vee}$. The elementary charges are the simple roots (co-roots), and their negatives. The interesting feature of this 'universal' Coulomb gas is that it presents a gas of particles of three charges: electric, magnetic and scalar. The first two interact with Coulomb-Coulomb interactions with particles of same charge type, or charge containing a root nearby on the Dynkin diagram. The scalar charges make the Coulomb gas unique as they interact such that like charges attract, and this introduces instability and exotic behaviour of the gas at different temperatures. The derivation of this exotic Coulomb gas for all gauge groups is the main result of this work and I hope that in the near future one will perform lattice Monte-Carlo simulations of this Coulomb gas, as done in [1] for the case of $SU(2)$, for all gauge groups.\\

As found in previous works [21], [23] the magnetic bions lead to mass gap for the dual photon fields $\sigma^a$ allowing for confinement of electric charges, and the neutral bions lead to a centre-stabilizing potential. The magnetic monopole-instantons do not lead to a mass gap as they contain fermionic zero modes and so were not considered as they cannot contribute to the vacuum structure and effective potential of the theory. It is noted that in studying the supersymmetric theory on a torus is that the theory is not as simple as the non-supersymmetric version, due to the presence of the adjoint scalar fields $\phi^a$, even though the GPY potential vanishes at zero temperature and partially cancels at $T>0$. Nonetheless, the dualities derived here are interesting and have led us to new phenomena and new ways of studying Yang-Mills theory at finite temperature.\\

Future directions of study include the following pursuits:\\

1. Lattice studies, as done in [1], can be done in this general gauge group setting, even if for particular gauge group such as $SU(3)$ or $G_2$, in both the dual Coulomb gas model or the XY-spin model, in order to gain better understanding of the phase transition as found in [2]. A first order phase transition is expected as opposed to the second order transition in the $SU(2)$ theory [1]. This can also lead to further study of the continuity conjecture as mentioned in [2], [7] by comparing phase transitions in pure thermal Yang-Mills to the quantum phase transitions in mass deformed super Yang-Mills. Comparison can be made to previous lattice studies and new studies in general gauge group may be possible as well. One must still obtain a dual spin model for SYM in other gauge groups than $SU(2)$ as future work before simulations of the dual spin model can be done. However, from the methods presented in [1], one may be able to do simulations of the dual double Coulomb gas for all gauge groups in a soon future work.\\

2. It has also been of recent interest to consider finite density QCD-like theories, in particular super Yang-Mills, and their phase transitions. There is a known sign problem due to finite chemical potential and so imaginary chemical potentials have been studied instead [3], [35]. This leads to a theory with twisted boundary conditions for the adjoint fermions along the compact direction (or directions). Computing the Callias index as a function of the twist angle leads to a twist-dependent index, which equals the usual answer, 2, at the centre-symmetric and supersymmetric vacuum. This recent work can be generalized to general gauge group and dependence on the boundary conditions is quite interesting.
\\

3. Mean field theory methods can be used for the XY spin models considered here, as well as related spin models in special cases. Although not exact, mean field theory can tell phase transitions and their orders, although at transition temperatures that are not always correct although within an order of magnitude. Studies of XY-models with symmetry breaking perturbations have been studied [44], [45] for different values of $p$ in the $\cos p\theta$-term and phase diagrams mapped out. It would be curious to implement a mean field theory that takes into account vortices and can verify known results, and produce new ones for other gauge groups not studied before. This would be interesting even for the case of zero scalar fields $\phi^a=0$. Cases with $\vec\phi\neq 0$ can be done as well in the mean field method. These cases are related to 'frustrated' XY models in the case that, on some lattices, bonds are ferromagnetic (like the scalar charged W's and neutral bions), while on others they are antiferromagnetic (like electric W's and magnetic bions). These competing interactions lead to 'frustration', that is a ground state that is degenerate and not at minimum possible energy without frustration. Models with competing F and AF interactions were studied in [44] and [46].
\\

4. Renormalization group equations and flow can be determined from the partition function (4.18). Special cases for $SU(2)$ and $SU(3)$ have been done with good success in [15] leading to known results of deconfinement, transition temperatures, and scaling parameters/critical exponents. In the cases of higher rank it was found that no fixed points appeared to exist for the RGEs and that the electric-magnetic duality no longer holds. It would be interesting to continue investigating the $SU(N)$ RGE cases and other groups of higher rank to see (possibly by going to higher order in the expansion) if there are fixed points and to find the nature of the critical points.
\\

5. General compactifications on toroidal spaces such as $\mathbb R^D\rightarrow \mathbb R^d\times (S^1)^{\times D-d}$ can be done in this generic case, although the applications or interests may not be immediate.\\

It is hoped that this work has provided a framework for future study with the goal of simulating the Coulomb gas derived as a main result of this work for all gauge groups. It is hoped as well that correct spin models for $SU(N)$ and other gauge groups, both for YM and SYM, can be done and lattice simulations of them performed, in order to compare to the Coulomb gas results for any gauge group and to determine the nature of the deconfining phase transition.

\acknowledgments

Special thanks to professor Erich Poppitz for fun projects and interesting research discussions, and suggestions for this paper. Special thanks as well to postdoc Mohamed Anber for helpful discussions and guidance.

\appendix

\section{Notes on Lie groups and Lie algebras}
\label{Liegroups}

For a sufficiently self-contained description of the mathematical constructs in our theory let us review Lie groups and Lie algebras. The familiar reader can skip to A.2 for the notation of roots and weights.

\subsection{Notes on general Lie theory}
\label{Liegeneral}

Let us begin by defining a Lie algebra and give its properties.\\
A {\it Lie algebra} $\mathfrak g$ is a vector space over a field $F$ (which we take here to be either real, $\mathbb R$, or complex, $\mathbb C$) with a binary operation (called the Lie bracket) $[\cdot,\cdot]\rightarrow\mathfrak g\times\mathfrak g\rightarrow\mathfrak g$ satisfying the basic properties:\\
(i) bilinearity: $[ax+by,cz+dw]=ac[x,y]+ad[x,w]+bc[y,z]+bd[y,w],\ \forall a,b,c,d\in F$ and $\forall x,y,z,w\in\mathfrak g$.\\
(ii) assymetry: $[x,y]=-[y,x],\ \forall x,y\in\mathfrak g$\\
(iii) Jacobi identity: $[x,[y,z]]+[z,[x,y]]+[y,[z,x]]=0,\ \forall x,y,z\in\mathfrak g$.\\
A Lie algebra is equipped with a basis of generators $\{T^a\}^r_{a=1}$ where $r=dim(\mathfrak g)$, and these satisfy the same relations above. The generators, forming a basis, have commutators which are linear combinations of generators, $[T^a,T^b]=f^{abc}T^c$, where the coefficients $f^{abc}$ are the structure constants of the algebra. In the fundamental representation this dimension is minimal and equal to the rank of its corresponding Lie group.\\

A Lie algebra is called {\it simple} if it is non-Abelian and has no non-zero proper ideals, and is {\it semi-simple} if it is non-Abelian and has no non-zero proper Abelian ideals. Hence a semi-simple Lie algebra $\mathfrak g$ can be written as a direct sum of simple Lie algebras $\mathfrak g_i$, $\mathfrak g=\oplus_{i=1}^n\mathfrak g_i$. We consider here just semi-simple Lie algebras.
\\

A Lie algebra varies depending on its representation. A {\it representation} $\mathcal R$ is given a map $\pi_{\mathcal R}:\mathfrak g\rightarrow\mathfrak{gl}(V)$, where $\mathfrak{gl}(V)$ is the enveloping associative Lie algebra of endomorphisms of a vector space $V$. The dimension of the representation $dim(\mathcal R)=dim(V)$ equals the dimension of the vector space $V$, if it is finite. For example, the fundamental representation has $dim(V)=rank(G)$. Also, in this paper, we use often the adjoint representation $ad:\mathfrak g\rightarrow\mathfrak{gl}(\mathfrak g)$ where the action is $ad(x)(y)=[x,y],\ \forall x,y\in\mathfrak g$.\\

A Lie group has a subgroup called the maximal torus $T\subset G$, whose elements commute with all other elements of the Lie group, and is topologically a torus $(S^1)^{\times r}$ where $r=dim(G)$, the topological dimension of the group. Its Lie algebra $\mathfrak t=Lie(T)$ is called the Cartan subalgebra of the Lie algebra and is of dimension $r$. Its generators $\{H^a\}^r_{a=1}$ with $[H^a,H^b]=0]$ form an $r$-dimensional subspace of $\mathfrak g$ and satisfy the normalization $tr(H^aH^b)=\delta^{ab}$ here.\\

The other generators of the Lie algebra can be represented by $dim(G)-r$ raising and lowering operators, $\{E_{\alpha}\}$ and $\{E_{-\alpha}=E_{\alpha}^{\dagger}\}$, which satisfy the relations
\begin{equation}
[H^i,E_{\alpha}]=\alpha^i E_{\alpha}
\end{equation}
$$[E_{\alpha},E_{-\alpha}]=\alpha_iH^i$$
$$[E_{\alpha},E_{\beta}]=N_{\alpha\beta\gamma}E_{\gamma},
$$
where the constants $N_{\alpha\beta\gamma}$ will not be needed later. The contravariant and covariant roots are related by the Cartan Killing form $g^{ij}=Tr[H^iH^j]$.
\\

A {\it Lie group}, as a reminder, is a group that is also a differentiable manifold, and hence has a differential structure or derivation (that satisfies the Leibnitz rule). In fact, its Lie algebra corresponding to it is the tangent space to the Lie group, specifically to its covering space $\tilde G$. Figure 2 shows all possible simply-connected, semi-simple Lie algebras and their Dynkin diagrams. For more definitions and detailed theory see [9].
\\


\subsection{The roots and the weights}
\label{rootsandweights}

One way to define the roots of a Lie group $G$ that will be useful later on is to consider it from the point of view of representations of its corresponding Lie algebra $\mathfrak g$. In general we define the root $\vec\alpha_i$ as an eigenvalue. In fact it is a function valued on $\mathfrak t=Lie T$, where $T$ is the maximal torus of $G$ $$\vec\alpha_i:\mathbb C\mathfrak t\rightarrow\mathbb C$$ with its eigenspace $E_{\alpha_i}\in \mathbb C\mathfrak g$ defined by
\begin{equation}
[H,E_{\alpha_i}]=\vec\alpha_i(H)E_{\alpha_i},
\end{equation}
where $H\in\mathbb C\mathfrak t$, the Cartan subalgebra of $\mathfrak g$.\\
We can see how this works for $SU(N)$. Beginning with $SU(2)$, we have Lie algebra $\mathfrak g=\mathfrak{sl_2}(\mathbb C)$. It is clear that $\mathfrak t= span\{\lambda \sigma_3=(^{\lambda\ \ 0}_{0\ \ -\lambda})\}_{\lambda\in\mathbb C}$ and that there are two root spaces, one with root the negative of the other: $E^{+}=span\{(^{0\ 1}_{0\ 0})\}$, $E^{-}=span\{(^{0\ 0}_{1\ 0})\}$. It is easy to check that the roots satisfying equation (A.2) are $\vec\alpha_{\pm}(H(\lambda))=\pm 2\lambda$.\\

This clearly generalizes to $SU(N)$ with Lie algebra $\mathfrak{sl}_N(\mathbb C)$. The maximal torus is just $T\approx\mathbb T^{N-1}\approx\{diag(e^{i\theta_j})_{j=1}^{N})|\prod^Ne^{i\theta_j}=1\}$. The Cartan subalgebra is the set of matrices with complex numbers $\lambda_j$ along the diagonal, accompanied by their negatives $-\lambda_j$, so as to make the trace vanish (this is in fact for the adjoint representation). The root spaces are just the span of each $E_{jk}$, the $N\times N$ matrix with a 1 in the $i,j$-th position and zeroes elsewhere. One easily checks that the roots obey $\alpha_{jk}(H(\lambda))=\lambda_j-\lambda_k$ along with their negatives from
\begin{equation}
[H(\lambda),E_{jk}]=(\lambda_j-\lambda_k)E_{jk}.
\end{equation}
This is why in the adjoint representation, the roots take on values given by differences of Wilson line eigenphases $\theta_j$.\\

We will need to know the weights in the adjoint representation, which I prove are the roots (the full set) of the Lie algebra. Let us describe representation theory in general a bit first.\\

A representation of a Lie algebra is a homomorphism from the Lie algebra $\mathfrak g$ into the endomorphism group of a certain vector space $V$,
$$\phi:\mathfrak g\rightarrow End(V),$$ and preserves the Lie bracket. The dimension of the representation is the dimension of the vector space $V$ underlying the representation. The dimension of the Lie algebra itself is the numbers of independent generators of $\mathfrak g$. In the fundamental representation the dimension of $\mathfrak g$ equals the dimension of $V$ and hence of the representation. The rank $r$, however, of a Lie algebra is the dimension of the Cartan subalgebra $\mathfrak h\subset\mathfrak g$. The Cartan subalgebra has a set of Abelian generators in the Cartan-Weyl basis $\{H^i\}_{i=1}^r$ satisfying $[H^i,H^j]=0$ and the roots, as mentioned before, satisfy eigenvalue-like expressions: $[H^i,E^{\alpha}]=\alpha_iE^{\alpha}$, and there are hence $r=rk(\mathfrak g)$ simple positive roots of $\mathfrak g$. The Lie algebra then decomposes as
$$\mathfrak g=\mathfrak h\oplus_{\alpha\in\Delta^+}\mathfrak g_{\alpha},$$ where $\mathfrak g_{\alpha}$ is the eigenspace, spanned by $E^{\alpha}$. We can also prove what the weights (eigenvalues of the $H^i$) are in fact the roots in the adjoint representation. Indeed,
\begin{equation}
\phi_{adj}(H^i)E^{\alpha}=ad_{H^i}E^{\alpha}\equiv [H^i,E^{\alpha}]=\alpha_iE^{\alpha},
\end{equation}
proving the claim. $\Box$\\

Below I will list the positive roots for each Lie algebra, but one more point to make is the role of the affine root in spaces with a compact direction. The Lie algebra with the affine root included is the Lie algebra of the loop group $LG$ of maps $\pi:S^1\rightarrow G$, $Lie(LG)=L\mathfrak g$. Similarly on spaces with multiple compact directions there are more roots to be added and the resulting algebra is the toroidal Lie algebra. The affine roots are included below.\\

For all semi-simple Lie groups (described below) a choice of simple positive roots is given as follows:\\

$A_{N+1}\approx SU(N)$:\\

This is the group of rotations about the origin in $\mathbb C^N$. It preserves the lengths of vectors. Their sets of simple roots are:
\begin{equation}
\{\alpha_i=e_i-e_{i+1}\}_{i=1}^N
\end{equation}
The affine root is $\alpha_0=-\sum_{i=1}^N\alpha_i=e_N-e_1$.\\

$B_N\approx Spin(2N+1)$:\\

This is the double cover of the orthogonal group $SO(2N+1)$, the rotation group in $\R^{2N+1}$. Here, the set of simple roots is given by
\begin{equation}
\{e_i-e_{i+1}\}_{1<i<N-1}\cup\{e_N\}
\end{equation}
The affine root is $-\alpha_0=e_1+e_2=\alpha_1+2\sum_{i=2}^N\alpha_i$.\\

$C_N \approx Sp(2N)$:\\

This is the group of $2N\times 2N$ matrices preserving the antisymmetric scalar product $J=(^{0\ \ \ 1_N}_{-1_N\ 0})$, so $M^TJM=J$ $\forall M\in Sp(2N)$. The simple roots are:
\begin{equation}
\{e_i-e_{i+1}\}_{1<i<N-1}\cup\{2e_N\}
\end{equation}
The affine root is $-\alpha_0=2e_1=\sum_{i=1}^{N-1}2\alpha_i+\alpha_N$.\\

$D_N\approx Spin(4N), Spin(4N+2)$ (N even, odd respectively):\\

These are the double covers of the orthogonal groups $SO(4N)$ and $SO(4N+2)$ respectively. The simple roots are:
\begin{equation}
\{e_i-e_{i+1}\}_{1<i<N-1}\cup\{e_{N-1}+e_N\}
\end{equation}
The affine root is $-\alpha_0=e_1+e_2=\alpha_1+2\sum_{i=2}^{N-2}\alpha_i+\alpha_{N-1}+\alpha_N$.\\

$E_6$:\\

This is the rank 6 exceptional Lie group of dimension 78. The simple roots are:
\begin{equation}
\{ (1,-1,0,0,0,0)
\end{equation}
$$(0,1,-1,0,0,0)$$
$$(0,0,1,-1,0,0)$$
$$(0,0,0,1,1,0)$$
$$-\frac{1}{2}(1,1,1,1,1,-\sqrt{3})$$
$$(0,0,0,1,-1,0)\}$$
The affine root is $-\alpha_0=e_1-e_8=\alpha_1+2\alpha_2+3\alpha_3+2\alpha_4+\alpha_5+2\alpha_6$. This is in the 8 dimensional basis and we note all vectors are orthogonal to $\sum_{i=1}^8e_i$ and to $e_1+e_8$ and so gauge fields are constrained by $\phi_1+\phi_8=\sum_{i=2}^7\phi_i=0$.\\

$E_7$:\\

This is the rank 7 exceptional group of dimension 133. The simple roots are:
\begin{equation}
\{ (1,-1,0,0,0,0,0)
\end{equation}
$$(0,1,-1,0,0,0,0)$$
$$(0,0,1,-1,0,0,0)$$
$$(0,0,0,1,-1,0,0)$$
$$(0,0,0,0,1,1,0)$$
$$-\frac{1}{2}(1,1,1,1,1,1,-\sqrt{2})$$
$$(0,0,0,0,1,-1,0)\}$$
The affine root is $-\alpha_0=e_2-e_1=2\alpha_1+3\alpha_2+4\alpha_3+3\alpha_4+2\alpha_5+\alpha_6+2\alpha_7$. The fields are constrained to live on the plane orthogonal to $\sum_{i=1}^8e_i=0$ in the 8 dimensional basis.\\

$E_8$:\\

This is the rank 8 exceptional group of dimension 248. The simple roots are:
\begin{equation}
\{ (1,-1,0,0,0,0,0,0)
\end{equation}
$$(0,1,-1,0,0,0,0,0)$$
$$(0,0,1,-1,0,0,0,0)$$
$$(0,0,0,1,-1,0,0,0)$$
$$(0,0,0,0,1,-1,0,0)$$
$$(0,0,0,0,0,1,1,0)$$
$$-\frac{1}{2}(1,1,1,1,1,1,1,1)$$
$$(0,0,0,0,0,1,-1,0)\}$$
The affine root is $-\alpha_0=e_1+e_2=2\alpha_1+3\alpha_2+4\alpha_3+5\alpha_4+6\alpha_5+4\alpha_6+2\alpha_7+3\alpha_8$.\\

$F_4$:\\

This rank 4 exceptional Lie group has dimension 52. Its simple roots are given by
\begin{equation}
\{ (0,1,-1,0)
\end{equation}
$$(0,0,1,-1)$$
$$(0,0,0,1)$$
$$-\frac{1}{2}(-1,1,1,1)\}$$
The affine root is $-\alpha_0=e_1+e_2=2\alpha_1+3\alpha_2+4\alpha_3+2\alpha_4$.\\

$G_2$: (embedded in 2D subspace of $\mathbb R^3$, the plane perpendicular to line $x+y+z=0$)\\

This is the rank 2 exceptional Lie group of dimension 14. Its simple roots are:
\begin{equation}
\{(0,1,-1),(1,-2,1)\}
\end{equation}
The affine root is $-\alpha_0=e_1+e_2-2e_3=2\alpha_1+3\alpha_2$. All vectors are orthogonal to $e_1+e_2+e_3=0$.\\

Note that the coefficients $k_i$ in the definition of the affine root $\vec\alpha_0=-\sum_{i=1}^rk_i\vec\alpha_i$ are called the Kac labels of the Lie algebra. The Coxeter number of the Lie algebra is $h(G)=\sum_{i=1}^rk_i+1$.\\

We will also need co-roots in our future defined as \begin{equation} \vec\alpha^{\vee}\equiv \frac{2}{\vec\alpha^2}\vec \alpha\in\Lambda^{\vee}_r, \end{equation} where they span the co-root lattice $\Lambda^{\vee}_r$, and the $\vec\alpha$'s are the $r=rank(G)$ simple roots given above and span the root lattice $\Lambda_r$.\\

{\bf The weights.}\\

We also need to get to know the weight system, with lattice $\Lambda_w$, and its co-weight lattice. The weight vectors $\vec w_i$ for a set of simple roots $\vec\alpha_i$ are defined via
\begin{equation}
\vec w_j\cdot \vec\alpha^{\vee}=\delta_{ij},
\end{equation}
and the co-weights are defined as were the co-roots:
\begin{equation}
\vec w^{\vee}\equiv \frac{2}{\vec w^2}\vec w\in\Lambda^{\vee}_w. \end{equation}
Since we are dealing with affine Lie algebras we need to define the affine co-root in terms of simple co-roots (the affine roots were given above, $\vec\alpha_0=-\sum_j^rk_j\vec\alpha_j$):
\begin{equation}
\vec\alpha_0^{\vee}=-\sum_j^rk_j^{\vee}\vec\alpha_j^{\vee},
\end{equation} and the dual Coxeter number is defined from the coefficients $c_2=\sum_{i=0}^rk_i^{\vee}$. For $\mathfrak g=\mathfrak{su}(r+1),\ c_2=r+1$ as all $k^{\vee}$'s are 1s, and all $\alpha$'s have norm $\sqrt{2}$. We will also soon need these data for $\mathfrak g_2$, where $c_2=4$ for $\{k_i^{\vee}\}=\{1,2,1\}$ and $\{\alpha_i^2\}=\{2,2,2/3\}$. For data such as these for all semi-simple Lie groups see [11]. Table 1 shows some such data including the Kac labels and (dual) Coxeter numbers. It is interesting to note that the Coxeter number $h$ of a group is the number of roots divided by the rank of the group. Figure 4 shows all semi-simple Lie algebras as (affine) Dynkin diagrams with the Kac labels included. As a reminder a Dynkin diagram is (for our purposes) a graph with single, double or triple lines connecting nodes, represented by simple roots. The multiplicity of the lines (edges) will not concern us, but are related to the length of roots represented by the nodes the edge connects. The affine Dynkin diagram, shown in Figure 4, contains the affine root $\vec\alpha_0$.\\

\begin{tabular}{l*{6}{c}r}
Group, G              & $r=rk(G)$ & $h$ & $c_2(G)$ & $[k_0^{\vee},\ldots,k_r^{\vee}]$ & $[k_0,\ldots,k_r]$ \\
\hline
$SU(N+1)$ & $N$ & $N+1$ & $N+1$ & $[1,1,\ldots,1]$ & $[1,1,\ldots,1]$ \\
$SO(2N+1)$ & $N$ & $2N$ & $2N-2$ & $[1,1,1,2,\ldots,2]$ & $[1,1,2,\ldots, 2]$ \\
$SO(2N)$ & $N$ & $2N$ & $2N-2$ & $[1,1,1,1,2,\ldots,2]$ & $[1,2,\ldots, 2,1]$\\
$Sp(2N)$ & $N$ & $2N-2$ & $N+1$ & $[1,1,\ldots,1]$ & $[1,1,2,\ldots,2,1,1]$\\
$G_2$ & 2 & $6$ & $4$ & $[1,1,2]$ & $[1,2,3]$\\
$F_4$ & 4 & $12$ & $9$ & $[1,1,2,3,2]$ & $[1,2,3,4,2]$\\
$E_6$ & 6 & $12$ & 12 & $[1,1,1,2,2,2,3]$ & $[1,1,2,3,2,1,2]$\\
$E_7$ & 7 & $18$ & 18 & $[1,1,2,2,2,3,3,4]$ & $[1,2,3,4,3,2,1,2]$\\
$E_8$ & 8 & $30$ & 30 & $[1,2,2,3,3,4,4,5,6]$ & $[1,2,3,4,5,6,4,2,3]$\\
\hline

\end{tabular}
\\

{\bf Table 4:} (Dual) Kac labels and dual Coxeter numbers for semi-simple Lie groups. Note that these in general differ from the Kac labels found from the table of simple roots above.\\

The weights of a Lie algebra in a given irrep $R$ represent the charges of particles possible for that irrep and hence are important Lie algebra data. The matrices $R(h)$ for any h in the Cartan subalgebra can be simultaneously diagonalized giving vectors $\vec w\in\mathfrak t^*$ of eigenvalues so that $\vec w\cdot h$ is an eigenvalue of $R(h)$. These vectors $\vec w$ belong to the set of weights of $R$ $\Delta_w^R$ and their integral span $\mathbb Z[\Delta_w^R]=\Lambda_w^R$ is called the weight lattice of $R$. The group lattice $\Gamma_G=\cup_R\Lambda_w^R$ is the union of irrep weight lattices. At the level of Lie group, the eigenvalues of irrep $R$ of an element $g\in T_G$, the maximal torus of $G$, are $\exp(2\pi i\vec w\cdot h)$. The periodicity of the maximal torus are given by shifts in the lattice of those $h$ such that $\vec w\cdot h\in\mathbb Z$. The dual lattice of co-weights is defined by the lattice of such $h$, $\Lambda_w^{R*}$. The smallest arising group lattice is called the root lattice $\Lambda_r$, whereas the largest is called the weight lattice $\Lambda_w$.
\\

For completeness I now present the weights of the adjoint representation (which are in fact the set of ALL roots as was shown above) for each Lie algebra. The number of weights is equal to the dimension of the representation minus the rank of the Lie algebra (the number of null weights of eigenvalue zero from the action of the Cartan generators.) \\

$A_{N+1}\approx SU(N)$:\\
There are $N^2+N$ adjoint weights in all, $N(N+1)/2$ being positive. All are of length $\sqrt{2}$ with 1 in one entry $i$, -1 in position $j$, and zeros elsewhere. We denote them as $\vec\alpha_{ij}^{\pm}$ where the superscript is positive if the root is. The positive roots are taken to be the ones with a +1 occuring in an earlier position than -1, i.e. $i<j$.\\

$B_N\approx Spin(2N+1)$:\\
There are $2N^2$ weights of two types: $\vec\alpha_{ij}^{\pm}$ which are all integer vectors of length $\sqrt{2}$, and $\beta_i^{\pm,B}$ which are all integer vectors of length 1. The positive weights are those with a +1 occurring before a -1 as usual.\\

$C_N\approx Sp(N)$:\\
In all there are $2N^2$ roots including the $\vec\alpha_{ij}^{\pm}$ above, and with $\beta_i^{\pm, C}=\pm 2e_i$.Positive roots are as before.\\

$D_N\approx Spin(4N),Spin(N+2)$:\\
Here all roots are all integer vectors of length $\sqrt{2}$. These include the $\vec\alpha_{ij}^{\pm}$ above, but also those with 2 entries both -1 or both +1, called $\vec\beta_{ij}^{\pm,D}$. There are $2N(N-1)$ in all.\\

$E_6$:\\
The adjoint weights include the $4\times (^5_2)$ permutations of the entries of the vectors $(\pm 1,\pm 1,0,0,0,0)$ keeping a zero in the last entry, plus the vectors of the form $\frac{1}{2}(\pm 1,\pm 1,\pm 1,\pm 1,\pm 1,\pm\sqrt{3})$ with an odd number of + signs. This gives a total of 72 weights.\\

$E_7$:\\
We have here $4\times (^6_2)$ permutations of $(\pm 1,\pm 1,0,0,0,0,0)$ keeping a zero in the last entry, plus the vectors of the form $\frac{1}{2}(\pm 1,\pm 1,\pm 1,\pm 1,\pm 1,\pm 1,\pm\sqrt{2})$ with an even number of + signs, plus the two vectors $(\vec 0,\pm\sqrt{2})$. This gives a total of 126 weights.\\

$E_8$:\\
We have 112 roots as permutations of $(\pm 1,\pm 1,0,0,0,0,0,0)$, plus the 128 vectors of the form $\frac{1}{2}(\pm 1,\pm 1,\pm 1,\pm 1,\pm 1,\pm 1,\pm 1,\pm 1)$ with an even number of - signs.\\

$F_4$:\\
We have here 48 roots: 24 as permutations of $(\pm 1,\pm 1,\vec 0)$ (call them type I), plus 8 roots as permutations of $(\pm 1,\vec 0)$ (type J), and 16 roots of the form $(\pm 1,\pm 1,\pm 1,\pm 1)/2$ (type K).\\

$G_2$:\\
Here there are 12 adjoint weights:
$$(1,-1,0),(2,-1,-1),(1,0,-1),(1,-2,1),(0,1,-1),(1,1,-2),$$ together with their negatives.\\

For more on Lie algebras, weights and representations, a great resource is [9].\\

{\bf Gauge cells and Weyl chambres.}\\

The Weyl group $W(\mathfrak g)$ is another group of gauge identifications on $\mathfrak t$, that acts as a group of linear transformations on $\mathfrak t$ that preserves the set of roots $\Delta_r$ (permutes them). It includes a Weyl reflection for each simple root $\alpha$ which acts on $h\in\mathfrak t^*$ by $\sigma_{\alpha}(h)=h-(h\cdot\vec\alpha^{\vee})\vec\alpha$. It acts on $\varphi\in\mathfrak t$ by $\sigma_{\alpha}(h)[\vec\varphi]=h\cdot\sigma_{\alpha}(\vec\varphi)$ and so $\sigma_{\alpha}(\vec\varphi)=\varphi-2(\vec\alpha\cdot\vec\varphi)\vec\alpha^{\vee}$ and is a reflection about the plane with normal vector $\vec\alpha$ passing through the origin. Allowing translations of the co-root lattice, the group of transformations is the semi-direct product $\hat W$ of $W$ and $\Gamma_r^{\vee}$. A fundamental domain or gauge cell (or affine Weyl chambre) $\hat{\mathfrak t}$ for $G$ is the quotient $\mathfrak t/\hat W$. A choice often used for the affine Weyl chambre is
$$\hat{\mathfrak t}=\{\vec\varphi\in\mathfrak t|0\leq\vec\alpha\cdot\vec\varphi,\ \forall\vec\alpha\in\Delta^s_r,\ -\vec\alpha_0\cdot\vec\varphi\leq 1\},$$
where $\Delta_r^s$ denotes the set of simple roots. This is the cell of interest here as it is the Cartan subalgebra modulo gauge equivalences. At points interior to $\hat{\mathfrak t}$ the unbroken gauge group is the maximal torus $U(1)^r$, while on the cell boundary, as mentioned previously, the gauge symmetry is enhanced due to elements being fixed by the gauge transformations $\hat W$, and the theory is no longer fully Abelianized. For explicit root systems and gauge cells see Appendix B of [4]. 

\section{Deriving the GPY potential and dual Coulomb gas}
\label{derivation}

\subsection{GPY effective potential derivation}

I present here a derivation of the GPY perturbative effective potential for general gauge group SYM on $\torus$, deriving (2.7).
\\

Recall that the determinant of an operator $\mathcal O$ is given by the product of its eigenvalues
\begin{equation}
Det\mathcal O=\prod_{\lambda}\lambda=\exp(\sum_{\lambda}\log\lambda),
\end{equation}
and using the zeta function
\begin{equation}
\zeta_{\mathcal O}(s)=\sum_{\lambda(\mathcal O)}\lambda^{-s}
\end{equation}
we find that the determinant and effective potential are respectively,
\begin{equation}
Det\mathcal O=\exp[-\zeta_{\mathcal O}'(0)],\ \ \ \ \ \ V_{bosonic}=-\zeta_{\mathcal O}'(0)/L\beta
\end{equation}
for the fermionic operator on $\mathcal R^2\times S^1_L\times S^1_{\beta}$. For the bosonic operator we need $[Det\mathcal O]^{-1}$ (it is -1 not -1/2 as there are two degrees of freedom/polarization for the gauge field) where our operator of interest is $\mathcal O=D_M^2$, so $V_{boson}=-\zeta_{D_M^2}'(0)/L\beta$. The eigenvalues of this operator are matrix-valued in the Lie group $G$
\begin{equation}
\lambda_{mn}=\vec k^2+(2\pi n/L+A_3^aT^a)^2+(2\pi m/\beta+A_0^aT^a)^2,
\end{equation}
where $\omega_n=2\pi n/L$ and $\Omega_m=2\pi m/\beta$ are the KK and Matsubara frequencies along the respective cycle of the torus $\mathbb T^2=S^1_L\times S^1_{\beta}$. By gauge invariance we can rotate our fields to have color components along the Cartan subalgebra $span\{H^a\}_{a=1}^r$. We also choose a gauge where the holonomies are constant (so $A_0,A_3$ are independent of $x_0$ and $x_3$). Our Wilson loops are then (using the vector notation to represent $r$-dimensional vectors in the Cartan subalgebra (the maximal torus) of $G$, and $\vec H=(H^1,\ldots,H^r)$)
$$\Omega_L=e^{iL\vec A_3\cdot \vec H},\ \ \ \ \ \ \Omega_{\beta}=e^{i\beta\vec A_0\cdot \vec H},$$
and we take the commutator of Wilson loops $[\Omega_L,\Omega_{\beta}]=0$ on the flat torus as it minimizes the effective potential. Writing the zeta function we must calculate (with trace in the adjoint representation in consideration)
\begin{equation}
\zeta(s)=\int \frac{d^2k}{(2\pi)^2}\sum_{(n,m)\in\mathbb Z^2}tr_{adj}[\vec k^2+(2\pi n/L+A_3^aT^a)^2+(2\pi m/\beta+A_0^aT^a)^2]^{-s}
\end{equation}
$$=\frac{1}{4\pi(s-1)}(\frac{\beta}{2\pi})^{2s-2}\sum_{(n,m)\in\mathbb Z^2}\sum_{\vec w\in\Delta_w}[(n\beta/L+\vec A_3\cdot\vec w\beta/2\pi)^2+(m+\vec A_0\cdot\vec w\beta/2\pi)^2]^{1-s}.
$$
In the second line we performed the integral over $d^2k=\pi d(k^2)$ and used the fact that the eigenvalues of the Cartan matrices form weight vectors, and that these are simply the roots of the Lie algebra of the Lie group $\mathfrak g=Lie(G)$ for the adjoint representation. See Appendix A for review on necessary concepts of Lie groups and algebras.\\

We need to find a low temperature expansion of this expression ($\beta>>L$) and so we use a useful identity (as used in [24])
\begin{equation}
\sum_{m\in\mathbb Z}\frac{1}{[(m+a)^2+c^2]^s}=\frac{\sqrt{\pi}}{\Gamma(s)}|c|^{1-2s}[\Gamma(s-1/2)+4\sum_{p=1}^{\infty}(\pi p|c|)^{s-1/2}\cos(2\pi pa)K_{s-1/2}(2\pi p|c|)].
\end{equation}
In our equation we take $a=\vec A_0\cdot\vec w\beta/2\pi$, $c=n\beta/L+\vec A_3\cdot\vec w\beta/2\pi$ and $s\rightarrow s-1$ to get (doing the sum over $m$)
\begin{equation}
\zeta(s)=\frac{1}{4\pi(s-1)}(\frac{\beta}{2\pi})^{2s-2}\frac{\sqrt{\pi}}{\Gamma(s-1)}\sum_{\vec w\in \Delta_w}\sum_{n\in\mathbb Z}|n\beta/L+\vec A_3\cdot\vec w\beta/2\pi|^{3-2s}[\Gamma(s-3/2) 
\end{equation}
$$+4\sum_{p=1}^{\infty}(\pi p|n\beta/L+\vec A_3\cdot\vec w\beta/2\pi|)^{s-3/2}\cos(2\pi p\vec A_0\cdot\vec w\beta/2\pi)K_{s-3/2}(2\pi p|n\beta/L+\vec A_3\cdot\vec w\beta/2\pi|)].$$
To find the effective perturbative bosonic potential we need only take the derivative of the overall divergent factor $\Gamma(s-1)^{-1}$ as taking derivatives of other terms will give zero at $s=0$. We note that $\frac{d}{ds}|_{s=0}\Gamma^{-1}(s-1)=\psi(-1)/\Gamma(-1)=+1$ ($\psi(z)$ is the logarithmic derivative of $\Gamma(z)$) and so we can safely set $s=0$. We notice that the first term is a Hurwitz zeta function which is related to a Bernouilli polynomial $B_4(z)=-\frac{3}{\pi^4}\sum_{k=1}^{\infty}\frac{\cos2\pi kz}{k^4}$ (times $(L/\beta)^{2s-3}$). The Hurwitz zeta function is related to Bernouilli polynomials in the following way
\begin{equation}
\zeta_H(s,z)=\sum_{n=0}^{\infty}\frac{1}{(n+z)^s}=B_{1-s}(z)/(s-1)=-\frac{\Gamma(2-s)}{(2\pi i)^{1-s}}\sum_{k\neq 0}\frac{e^{2\pi i kz}}{k^{1-s}}.
\end{equation}
so that $\zeta_H(-3,\vec A_3\cdot\vec wL/2\pi)=-\frac{1}{4}B_4(\vec A_3\cdot\vec wL/2\pi)$ and we note that the first term is exactly that obtained previously on $\mathbb R^3\times S^1_L$ as required at zero temperature [1], [39]: $-\frac{\pi^2}{12L^4}B_4(\vec A_3\cdot\vec wL/2\pi)$. This cancels the fermionic contribution so we are left with the remaining terms. The second term in (B.7) can be simplified using $K_{-3/2}(z)=\sqrt{\frac{2}{z}}(1+z^{-1})e^{-z}$. Collecting terms together we can get an exact expression using polylogarithms $Li_n(z)=\sum_{k=1}^{\infty}\frac{z^k}{k^n}$ and that $\Gamma(-3/2)=4\sqrt{\pi}/3$
\begin{equation}
V_{eff}^{pert., bosonic}=-\sum_{\vec w\in\Delta_w}[\frac{\pi^2}{12L^4}B_4(\vec A_3\cdot\vec w L/2\pi)+(L\beta)^{-2}\sum_{m=-\infty}^{\infty} 
\end{equation}
$$[|m+\vec A_3\cdot\vec w L/2\pi|Li_2(\exp(i\vec A_0\cdot\vec w\beta-\frac{\beta}{L}(\vec A_3\cdot\vec w L+2\pi m)))+(2\pi)^{-1}Li_3(\exp(i\vec A_0\cdot\vec w\beta-\frac{\beta}{L}(\vec A_3\cdot\vec w L+2\pi m)))+ h.c.]$$
We can alternatively write for the bosonic contribution:
\begin{equation}
V_{eff}^{boson}=\sum_{\vec w\in\Delta_w}[-\frac{2}{\pi^2L^4}\sum_{p=1}^{\infty}\frac{\cos pL\vec A_3\cdot \vec w}{p^4}-\sum_{n\in\mathbb Z}\sum_{p=1}^{\infty}\frac{e^{-2\pi p|n\beta/L+\beta\vec A_3\cdot\vec w/2\pi|}}{\pi\beta^3Lp^3}(1+
\end{equation}
$$2\pi p|n\beta/L+\beta\vec A_3\cdot\vec w/2\pi|)\cos(p\beta\vec A_0\cdot\vec w)].$$
To add the fermionic contribution we need the determinant $Det\slashed D$, which can be computed from $\slashed D^2=D_M^2-\sigma_{MN}F_{MN}/2$, which is the same as our bosonic operator as $F_{MN}=0$ along the holonomy's directions, and use $\log\slashed D=\log\slashed D^2/2$. We subtract the same result above but with $2m\rightarrow 2m+1$ due to the anti-periodic boundary conditions in the thermal direction. We get:
\begin{equation}
V_{eff}^{fermionic}=\sum_{\vec w\in\Delta_w}[\frac{2}{\pi^2L^4}\sum_{p=1}^{\infty}\frac{\cos pL\vec A_3\cdot \vec w}{p^4}+\sum_{n\in\mathbb Z}\sum_{p=1}^{\infty}(-1)^p\frac{e^{-2\pi p|n\beta/L+\beta\vec A_3\cdot\vec w/2\pi|}}{\pi\beta^3Lp^3}(1+
\end{equation}
$$2\pi p|n\beta/L+\beta\vec A_3\cdot\vec w/2\pi|)\cos(p\beta\vec A_0\cdot\vec w)].$$
Combining everything together we get
\begin{equation}
V_{eff}^{pert.}=\sum_{p=1}^{\infty}[(-1)^p-1]\sum_{\vec w\in\Delta_w}\sum_{n\in\mathbb Z}\frac{e^{-2\pi p|n\beta/L+\beta\vec A_3\cdot\vec w/2\pi|}}{\pi\beta^3Lp^3}(1+2\pi p|n\beta/L+\beta\vec A_3\cdot\vec w/2\pi|)\cos(p\beta\vec A_0\cdot\vec w).
\end{equation}

We now look at the low temperature contribution and consider just the $p=1$ term as other terms are suppressed by higher powers of the Boltzmann factor $e^{-m_W/T}=e^{-\beta/L}$. The result is
\begin{equation}
V_{eff}^{pert., low T}(\vec A_0,\vec A_3)\approx -2\sum_{\vec w\in \Delta_w}\sum_{n\in\mathbb Z}\frac{e^{-2\pi |n\beta/L+\beta\vec A_3\cdot\vec w/2\pi|}}{\pi\beta^3L}(1+2\pi |n\beta/L+\beta\vec A_3\cdot\vec w/2\pi|)\cos(\beta\vec A_0\cdot\vec w).
\end{equation}

\subsection{Dual 'double' Coulomb gas derivation details}

Here I present the series of dualities used to arrive at the dual Coulomb gas partition function (4.12).
To evaluate the path integral over the gauge fields in (4.7) we can go one step further and enhance the $U(1)^r$ symmetry of the fields $\sigma^a$ to $U(1)^{2r}$ by introducing another set of scalar fields $\lambda^a$ and new fields $\Phi^a_{\mu}$ and $\vartheta^a$ [6], which transform under the two $U(1)^r$ symmetries as $$\vartheta^a\rightarrow\vartheta^a-\lambda^a,\ \ \ \ \Phi^a_{\mu}\rightarrow \Phi^a_{\mu}+\partial_{\mu}\lambda^a, \ \ \ \ A_{\mu}^a\rightarrow A_{\mu}^a+\partial_{\mu}\sigma^a.$$ The path integral over the gauge fields becomes
\begin{equation}
\int\mathcal D \vec A_{\mu}e^{-\int d^3x(\vec F_{\mu\nu})^2/4g_3^2}=\int\mathcal D\vec A_{\mu}\mathcal D\vec\Phi_{\nu}\mathcal D\vec\vartheta\exp[-\int d^3x(g_3^2(\partial_{\mu}\vec\vartheta+\vec\Phi_{\mu})^2/2+i\sum_{\vec w\in\Delta_w^{\mathcal R}}\epsilon_{\mu\nu\rho}\partial_{\mu}\vec A_{\nu}\cdot\vec w\vec\Phi_{\rho}\cdot\vec w)],
\end{equation}
where $g_3$ is the effective three dimensional coupling. Taking $\vec\vartheta=0$ (unitary gauge) and varying the new action with respect to $\vec\Phi_{\mu}$ gives $\vec\Phi_{\mu}=-i\epsilon_{\mu\nu\rho}\partial_{\nu}\vec A_{\rho}/g_3^2$, which once substituted in the Lagrangian in (B.14) yields the original Lagrangian $(\vec F_{\mu\nu})^2/4g_3^2$. We now evaluate (I used $\sum_{\vec w\in\Delta_w^{\mathcal R}}\vec X\cdot \vec w\vec Y\cdot\vec w=\vec X\cdot\vec Y$)
\begin{equation}
\int\mathcal D\vec A_{\mu}^{ph}\int\mathcal D\vec\Phi_{\nu}\exp[-\int d^{2+1}x[\frac{1}{4g_3^2}(\vec F_{\mu\nu}^{bion})^2+\frac{1}{2g_3^2}\vec F_{\mu\nu}^{bion}\cdot\vec F_{\mu\nu}^{ph}+\frac{g_3^2}{2}\vec\Phi_{\mu}^2+i\epsilon_{\mu\nu\rho}\partial_{\mu}\vec A_{\nu}^{ph}\cdot\vec \Phi_{\rho}]]
\end{equation}
$$\times \exp[-\int d^{2+1}xi\sum_A\sum_{a=0}^rq_A\vec\alpha_a\cdot\vec A_0^{ph}(\vec x,x_0)\delta(\vec x-\vec x_A^a)].
$$

Using the new action in (B.14) allows us to work with $S_{aux}$ given below where we substituted for $\vec F_{\mu\nu}^2/4g_3^2$:
\begin{equation}
S_{aux}=\int d^3x[\frac{1}{4g_3^2}(\vec F_{\mu\nu}^{bion})^2+\frac{1}{2g_3^2}\vec F_{\mu\nu}^{bion}\cdot\vec F^{\mu\nu,ph}+\half g_3^2\vec\Phi_{\mu}^2+i\epsilon_{\mu\nu\lambda}\partial_{\mu}\vec A_{\nu}^{ph}\cdot\vec\Phi_{\lambda}-i\sum_{a=0}^r\sum_Aq_A^a\vec\alpha_a\cdot\vec A_{\mu}^{ph}(\vec x, x_0)\delta^2(\vec x-\vec x_A^a)].
\end{equation}
Varying this with respect to $\vec A_{\mu}^{ph}$ gives $i\epsilon_{\mu\nu\lambda}\partial_{\mu}\vec\Phi_{\lambda}+\partial_{\mu}\vec F_{\mu\nu}^{bion}/g_3^2=-i\sum_{a=0}^r\sum_Aq_A^a\vec\alpha_a\delta^2(\vec x-\vec x_A)\delta_{0\nu}$ which has solution $\vec\Phi_{\mu}=i\vec B_{\mu}/g_3^2+\vec b_{\mu}$, where $\vec B_{\mu}=\epsilon_{\mu\nu\lambda}\vec F_{\nu\lambda}^{bion}/2$ is the magnetic field due to the background of magnetic bions and $\vec b_{\mu}$ splits into its divergence and curl free parts: $\vec b_{\mu}=\partial_{\mu}\vec\sigma+\epsilon_{\mu\nu\lambda}\partial_{\nu}\vec C_{\lambda}$ with $\partial_{\nu}\vec C_{\nu}=0$. Putting $\vec b_{\mu}$ back in $S_{aux}$ shows that $\vec\sigma$ drops out and the equation of motion for $\vec C_{\mu}$ is $\nabla^2\vec C_{\mu}=-\sum_{a=0}^r\sum_Aq_A^a\vec\alpha_a\delta^2(\vec x-\vec x^a_A)\delta_{0\mu}$. Introducing a Green's function $G(x)$ on $\mathbb R^2\times S^1_{\beta}$ satisfying $\nabla^2 G(\vec x-\vec x',x_0-x_0')=-\delta^2(\vec x-\vec x')\delta(x_0-x_0')$ with solution $G(\vec x-\vec x',x_0-x_0')=\frac{1}{4\pi}\sum_{n\in\mathbb Z}\frac{1}{\sqrt{(\vec x-\vec x')^2+(x_0-x_0'+n\beta)^2}}$ gives the solution for $\vec C_{\nu}$,
\begin{equation}
\vec C_{\mu}=\sum_{a=0}^r\int d^3{x}'\sum_Aq_A^a\vec\alpha_a\delta_{0\mu}\delta^2(\vec {x}'-\vec x_{A}^a)G(\vec x-\vec {x^a}',x_{0}-{x_{0}^a}')=
\end{equation}
$$=\frac{1}{4\pi}\sum_{a=0}^r\sum_Aq_A\vec\alpha_a\delta_{0\mu}\oint_{S^1_{\beta}}dx_0\sum_{n\in\mathbb Z}\frac{1}{\sqrt{(\vec x-\vec x_{A}^a)^2+(x_{0}-x_{0A}^a+n\beta)^2}}=-\frac{\delta_{0\mu}}{2\pi}\sum_{a=0}^r\sum_Aq_A\vec\alpha_a\log|\vec x-\vec x_{A}^a|.
$$
The solution for $\vec\Phi_{\mu}=i\vec B_{\mu}/g_3^2+\vec b_{\mu}+\vec K_{\mu}$ can also be found where a term $\vec K_{\mu}=-\frac{\epsilon_{0\mu\nu}}{2\pi}\sum_{a=0}^r\sum_Aq_A^a\vec\alpha_a\partial_{\mu}\log|\vec x-\vec x_A^a|$. Substituting this into $S_{aux}$ and integrating by parts yields
\begin{equation}
S_{aux}=\int d^3x\half g_3^2[(\partial_{\mu}\vec\sigma)^2+\vec K_{\mu}^2]-i\vec\sigma\cdot\partial_{\mu}\vec B_{\mu}+i\vec B_{\mu}\cdot\vec K_{\mu}-g_3^2\vec\sigma\cdot\partial_{\mu}\vec K_{\mu},
\end{equation}
where the last term is zero from the asymmetry of the $\epsilon_{\mu\nu 0}$ and $\partial_{\mu}\vec B_{\mu}=4\pi\sum_{a=0}^r\sum_iq_i^a\vec\alpha_a^{\vee}\delta^3(x-x_i^a)$ from Gauss' law. We can then write $\vec B_{\mu}=\sum_{a=0}^r\sum_iq_i^a\vec\alpha_a^{\vee}(\frac{(x-x_i^a)_{\mu}}{|x-x_i^a|^3})^{(p)}$ where $(p)$ denotes the periodicity enforced along the $S^1_{\beta}$. The next term $i\vec B_{\mu}\cdot\vec K_{\mu}$ in $S_{aux}$ can be evaluated as well and is seen to be zero:
$$\int d^3x\vec B_{\mu}\cdot\vec K_{\mu}=-\int d^2x\sum_{a,b=0}^r\sum_{iA}q_i^aq_A^b\vec\alpha_a^{\vee}\cdot\vec\alpha_b\oint_0^{\beta}\sum_{n\in\mathbb Z}\frac{\epsilon_{kl}(x-x_i^a)_k(x-x_A^b)_l}{|\vec x-\vec x_A^b|^2|(\vec x-\vec x_i^a)^2+(x_0-x_{0i}^a+n\beta)^2|^{3/2}}$$
$$=-2\int d^2x\sum_{a,b=0}^r\sum_{iA}q_i^aq_A^b\vec\alpha_a^{\vee}\cdot\vec\alpha_b\frac{\epsilon_{kl}(x-x_i^a)_k(x-x_A^a)_l}{|\vec x-\vec x_A^b|^2|\vec x-\vec x_i^a|^2}=0$$ by symmetric integration under asymmetric $\epsilon_{ij}$.
\\
What remains is the $\vec K_{\mu}^2$ term in $S_{aux}$, and we will see this gives rise to the Coulomb interactions of the W-bosons.
$$\int d^3x\vec K_{\mu}^2=\frac{1}{(2\pi)^2T}\sum_{a,b=0}^r\sum_{AB}q_A^aq_B^b\vec\alpha_a\cdot\vec\alpha_b\int d^2x\partial_i\log|\vec x-\vec x_A^a|\partial_i\log|\vec x-\vec x_B^a|$$ $$=\frac{1}{(2\pi)^2T}\sum_{a,b=0}^r\sum_{AB}q_A^aq_B^b\vec\alpha_a\cdot\vec\alpha_b\int d^2x\frac{(x-x_A)_k(x-x_B)_k}{|\vec x-\vec x_A^a|^2|\vec x-\vec x_B^b|^2}=-\frac{1}{2\pi T}\sum_{a,b=0}^r\sum_{AB}q_A^aq_B^b\vec\alpha_a\cdot\vec\alpha_b\log(T|\vec x_A^a-\vec x_B^b|).$$

Further, varying $S_{aux}$ with respect to $\vec\sigma$ gives $g_3^2\nabla^2\vec\sigma=-4\pi i\sum_{a=0}^r\sum_iq_i^a\delta^3(x-x_i^a)$, which has solution $\vec\sigma=\frac{4\pi i}{g_3^2}\sum_{a=0}^r\sum_iq_i^a\vec\alpha_a^{\vee} G(\vec x-\vec x_i^a,x_0-x_{0i}^a)$. Putting this back into $S_{aux}$ gives $S_{aux}=\frac{8\pi}{g_3^2}\sum_{a,b=0}^r\sum_{ij}q_i^aq_j^b\vec\alpha_a^{\vee}\cdot\vec\alpha_b^{\vee}G(\vec x_j^b-\vec x_i^a,x_{0j}^b-x_{0i}^a)-\frac{g_3^2}{4\pi T}\sum_{a,b=0}^r\sum_{AB}q_A^aq_B^b\vec\alpha_a^{\vee}\cdot\vec\alpha_b^{\vee}\log(T|\vec x_A^a-\vec x_B^b|)$. This gives us our final partition function for the double Coulomb gas as given in (4.) once the $\vec\phi$ terms are restored (reminding ourselves that the W-boson fugacity depends on $\phi$ once we turn it on). Putting everything together we get the final result
\begin{equation}
\mathcal Z_{grand}=\sum_{N_{b\pm},q_i^a=\pm}\sum_{N_{W\pm},q_A^a=\pm}\frac{\xi_b^{N_{b+}+N_{b-}}}{N_{b+}!N_{b-}!}\prod_i^{N_{b+}+N_{b-}}\int\prod_{a=0}^r d^3x_i^a\frac{(T\xi_W(\phi))^{N_{W+}+N_{W-}}}{N_{W+}!N_{W-}!}\prod_A^{N_{W+}+N_{W-}}
\end{equation}
$$\int \prod_{a=0}^rd^3x_A^a\int\mathcal D\vec\phi\exp[\sum_{a,b=0}^r(\frac{32\pi LT}{g^2}\sum_{i>j}q_i^aq_j^b\vec\alpha_a^{\vee}\cdot\vec\alpha_b^{\vee}\log|\vec x_i^a-\vec x_j^b|+\frac{g^2}{2\pi LT}\sum_{A>B}q_A^aq_B^b\vec\alpha_a\cdot\vec\alpha_b\log|\vec x_A^a-\vec x_B^b|+$$
$$4i\sum_{i,A}q_i^aq_A^b\vec\alpha_a^{\vee}\cdot\vec\alpha_b\Theta(\vec x_A^b-\vec x_i^a))+\int_{\mathbb R^2}[\half\frac{g^2}{(4\pi)^2L}(\partial_{\mu}\vec\phi)^2+V^0_{bion}\sum_{a=0}^r(k_a^{\vee}\vec\alpha_a^{\vee})^2\exp(-\frac{8\pi}{g^2}\vec\alpha_a^{\vee}\cdot\vec\phi)]],$$
which is valid for all $T$ with $0\leq T<M_W$, and I used $g_3=g/L$ and the long-distance property of the Green's function where it behaves like a logarithm. This concludes the steps to derive the dual double Coulomb gas (4.12). 

\section{Monopole solutions for all simple groups}
\label{monopoles}

I present here monopole solutions used in section 3 and give their actions. Let me review the $SU(2)$ case first, where we have one BPS and one KK solution on $\mathbb R^3\times S^1$. The solution exists due to the symmetry breaking $SU(2)\rightarrow U(1)$ so that $\pi_2(SU(2)/U(1))\approx\pi_1(U(1))\approx\mathbb Z$.
\\

In general recall the Euclidean action of pure Yang-Mills splits into electric and magnetic field energy components.
\begin{equation}
S=\frac{1}{2g^2}\int_{\mathbb R^3\times S^1}tr[F_{MN}^aT^aF^{bMN}T^b]=\frac{L}{g^2}\int d^3x tr[B^a_{\mu}T^aB_{\mu}^bT^b+E^a_{\mu}T^aE_{\mu}^bT^b],
\end{equation}
where $B_{\mu}^a=\epsilon_{\mu\nu\lambda}F_{\nu\lambda}^a$ and $E_{\mu}^a=D_{\mu}A_3^a$, and the generators are taken in the fundamental representation. The action (C.1) can be rewritten as
\begin{equation}
S=\frac{L}{g^2}\int d^3xtr[(B_{\mu}^a\mp D_{\mu}A_3^a)T^a(B_{\mu}^b\pm D_{\mu}A_3^b)T^b]\pm 2tr[B_{\mu}^aT^aD_{\mu}A_3^bT^b].
\end{equation}
The last term can be integrated by parts and the equation of motion $D_{\mu}B_{\mu}^a=0$ so we have the action as
\begin{equation}
S=\frac{L}{g^2}\int d^3xtr[(B_{\mu}^a\mp D_{\mu}A_3^a)T^a(B_{\mu}^b\pm D_{\mu}A_3^b)T^b]
\end{equation}
$$\pm 2\frac{L}{g^2}\int_{S^2_{\infty}}d^2\Sigma_{\mu}tr[B_{\mu}^aT^aA_3^bT^b].$$
Our monopoles satisfy self-dual or anti self-dual equations $B_{\mu}^a=\pm D_{\mu}A_3^a$ (equivalently, $F_{MN}=\pm\tilde F_{MN}$) which simplifies the action to (using $tr[T^aT^b]=\delta^{ab}/2$)
\begin{equation}
S=\pm\frac{L}{g^2}\int_{S^2_{\infty}}d^2\Sigma_{\mu}B_{\mu}^aA_3^a].
\end{equation}

The instanton number satisfies 
\begin{equation}
\mathcal K=(16\pi)^{-1}\int_{\mathbb R^3\times S^1}tr[F_{MN}^aT^a\tilde F_{MN}^bT^b]=\pm\frac{g^2}{8\pi^2}S,
\end{equation}
where the last equality holds for self dual and anti self dual solutions respectively. Setting $\mathcal K=1$ gives the usual monopole action as required: $S=8\pi^2/g^2$.

\subsection{$SU(2)$ monopole solution}
\label{SU(2)monopole}

For $SU(2)$ I present the solution as given in [2]. We take as generators $T^a=\tau^a/2$ where $\tau^a$ are the Pauli matrices, and satisfy $tr[T^aT^b]=\delta^{ab}/2$. The action is
\begin{equation}
S=\frac{1}{4g^2}\int_{\mathbb R^3\times S^1}F^a_{MN}F^{aMN}.
\end{equation}
In the hedgehog gauge the monopole solution is given by
\begin{equation}
A_{\mu}=A_{\mu}^c\tau^c=\epsilon_{\mu\nu c}\frac{x_{\nu}}{|x|^2}(1-\frac{v|x|}{\sinh(v|x|)})\tau^c,
\end{equation}
$$A_3=\Psi^c\tau^c=\frac{x_c}{|x|^2}(v|x|\coth(v|x|)+1)\tau^c,$$
where $v=\langle A_3\rangle$. Putting these solutions (C.7) into the action (C.6) gives $S=4\pi vL/g^2$ so that $v=2\pi/L$ gives the usual monopole instanton action $8\pi^2/g^2$.\\

The magnetic field's asymptotics are found to be (in the string/singular gauge)
\begin{equation}
B_{\mu}=\frac{1}{2}\epsilon_{\mu\nu\lambda}F_{\nu\lambda}\rightarrow_{|x|\rightarrow\infty}-\frac{x_{\mu}}{2|x|^3}\tau^3.
\end{equation}

\subsection{Monopole solutions for arbitrary gauge group $G$}
\label{monopolegroups}

To find the monopole solutions for general gauge group we can embed the $SU(2)$ solution into $G$, $SU(2)\subset G$, for each simple co-root $\vec\alpha_i^{\vee}$. The KK monopole solution will be given later to give $r+1$ monopole solutions as is consistent with the symmetry breaking $G\rightarrow U(1)^r$ (for the case of full Abelianization), which presents $r$ BPS solutions as $\pi_2(G/U(1)^r)\approx\pi_1(U(1)^r)\approx\mathbb Z^r$ (since $\pi_2(G)\approx 0$ for covering spaces of Lie groups $\tilde G$, which we consider here as they allow all representations, in particular spin representations). The KK solution arises due to the compact direction and will be given later associated to the affine co-root $\vec\alpha_0^{\vee}$.
\\

The $SU(2)$ embedding into $G$ for each simple root $\vec\alpha_i$ is given by
\begin{equation}
t^1=\frac{1}{\sqrt{2\vec\alpha_i^2}}(E_{\alpha_i}+E_{-\alpha_i}),\ t^2=\frac{1}{\sqrt{2i\vec\alpha_i^2}}(E_{\alpha_i}-E_{-\alpha_i}),\ t^3=\frac{1}{2}\vec\alpha_i^{\vee}\cdot\vec H,
\end{equation}
which obey the $SU(2)$ algebra commutation relations $[t^a,t^b]=i\epsilon^{abc}t^c$. The solutions for the gauge field are the same as (C.7) but with
\begin{equation}
A_3=\Psi^c\tau^c+(\vec\phi-\frac{1}{2}\vec\alpha_i^{\vee}v)\cdot\vec H,
\end{equation}
where $\vec\phi$ determines the asymptotics of the gauge field $v=\vec A_3\cdot\vec\alpha_i=\vec\alpha_i\cdot\vec\phi/L$. The solution $A_3$ is as is to guarantee these asymptotics since (in string gauge)
$$\Psi^c\tau^c|_{|x|\rightarrow\infty}=\frac{x^c}{|x|}t^c\frac{\vec\alpha_i\cdot\vec\phi}{L}=t^3\frac{\vec\alpha_i\cdot\vec\phi}{L}=\frac{1}{2}\frac{\vec\alpha_i\cdot\vec\phi}{L}\vec\alpha_i^{\vee}\cdot\vec H.$$
The BPS magnetic monopole's magnetic field's asymptotics are given by
\begin{equation}
B_{\mu}^{\alpha_i}=-\frac{x_{\mu}}{|x|^3}\frac{\vec\alpha_i^{\vee}\cdot\vec H}{2},\ i=1,\ldots, r.
\end{equation}
Its action and instanton number are given by
\begin{equation}
S^{\alpha_i}=\frac{4\pi}{g^2}\vec\alpha_i^{\vee}\cdot\vec\phi,\ \ \ \mathcal K^{\alpha_i}=\frac{\vec\alpha_i^{\vee}\cdot\vec\phi}{2\pi},
\end{equation}
respectively.\\

The other solution mentioned before, the KK monopole, can be found by a Weyl reflection as in [14]. Its asymptotic magnetic field is 
\begin{equation}
B_{\mu}^{\alpha_0}=-\frac{x_{\mu}}{|x|^3}\frac{\vec\alpha_0^{\vee}\cdot\vec H}{2}.
\end{equation}
Note that it has negative magnetic charge. (Also, since $\vec\alpha_0^{\vee}=-\sum_{i=1}^rk_i^{\vee}\vec\alpha_i^{\vee}$ an instanton can be formed from the collection of $2c_2(G)$ monopoles.) Its action and monopole number are found to be [2]
\begin{equation}
S^{\alpha_0}=\frac{4\pi}{g^2}(2\pi+\vec\alpha_0^{\vee}\cdot\vec\phi),\ \ \ \mathcal K^{\alpha_0}=\frac{2\pi+\vec\alpha_0^{\vee}\cdot\vec\phi}{2\pi}.
\end{equation}

\end{document}